\documentclass{pasj02}
\Received{$\langle$reception date$\rangle$}
\Accepted{$\langle$acception date$\rangle$}
\Published{$\langle$publication date$\rangle$}

\usepackage{physics,siunitx,amsmath,threeparttable}
\usepackage[switch,mathlines]{lineno} 
\usepackage{comment,subcaption,tabularx}
\begin{document}

\title{XRISM spectroscopy on orbital modulation of Fe Ly$\boldsymbol{\alpha}$ lines in Cygnus X-3}
\author{Daiki \textsc{Miura}\altaffilmark{1,2,}\footnotemark[*]\orcid{0009-0009-0439-1866}, Hiroya \textsc{Yamaguchi}\altaffilmark{2,1,3}\orcid{0000-0002-5092-6085}, Ralf \textsc{Ballhausen}\altaffilmark{4,5,6}\orcid{0000-0002-1118-8470}, Timothy \textsc{Kallman}\altaffilmark{5}, Teruaki \textsc{Enoto}\altaffilmark{7}\orcid{0000-0003-1244-3100}, Shinya \textsc{Yamada}\altaffilmark{8}\orcid{0000-0003-4808-893X}, Tomohiro \textsc{Hakamata}\altaffilmark{9}, Ryota \textsc{Tomaru}\altaffilmark{9}\orcid{0000-0002-6797-2539}, Hirokazu \textsc{Odaka}\altaffilmark{9}, Natalie \textsc{Hell}\altaffilmark{10}\orcid{0000-0003-3057-1536}, Hiroshi \textsc{Nakajima}\altaffilmark{11}\orcid{0000-0001-6988-3938}, Shin \textsc{Watanabe}\altaffilmark{2}\orcid{0000-0003-0441-7404}, Tasuku \textsc{Hayashi}\altaffilmark{8}, Shunji \textsc{Kitamoto}\altaffilmark{8}\orcid{0000-0001-8948-7983}, Kazutaka \textsc{Yamaoka}\altaffilmark{12}\orcid{0000-0003-3841-0980}, Jon M. \textsc{Miller}\altaffilmark{13}, Keigo \textsc{Okabe}\altaffilmark{3}\orcid{0009-0004-9550-3643}, Itsuki \textsc{Maruzuka}\altaffilmark{3}\orcid{0009-0000-8402-9211}, Karri \textsc{Koljonen}\altaffilmark{14}\orcid{0000-0002-9677-1533} and Mike \textsc{McCollough}\altaffilmark{15}\orcid{0000-0002-8384-3374}}
\altaffiltext{1}{Department of Physics, Graduate School of Science, The University of Tokyo, 7-3-1 Hongo, Bunkyo-ku, Tokyo 113-0033, Japan}
\altaffiltext{2}{Institute of Space and Astronautical Science (ISAS), Japan Aerospace Exploration Agency (JAXA), 3-1-1 Yoshinodai, Chuo-ku, Sagamihara, Kanagawa 252-5210,
Japan}
\altaffiltext{3}{Department of Science and Engineering, Graduate School of Science and Engineering, Aoyama Gakuin University, 5-10-1, Fuchinobe, Sagamihara 252-5258, Japan}
\altaffiltext{4}{University of Maryland College Park, Department of Astronomy, College Park, MD 20742, USA}
\altaffiltext{5}{NASA Goddard Space Flight Center, 8800 Greenbelt Rd, Greenbelt, MD 20771, USA}
\altaffiltext{6}{Center for Research and Exploration in Space Science and Technology, NASA / GSFC (CRESST II), Greenbelt, MD 20771, USA}
\altaffiltext{7}{Department of Physics, Kyoto University, Kitashirakawa-Oiwake-cho, Sakyo-ku, Kyoto, 606-8502, Japan}
\altaffiltext{8}{Department of Physics, Rikkyo University, Toshima-Ku, Tokyo, 171-8501, Japan}
\email{dikmur0611@g.ecc.u-tokyo.ac.jp}
\altaffiltext{9}{Department of Earth and Space Science, Graduate School of Science, Osaka University, 1-1 Machikaneyama, Toyonaka, Osaka 560-0043, Japan}
\altaffiltext{10}{Lawrence Livermore National Laboratory, 7000 East Avenue, Livermore, CA 94550, USA}
\altaffiltext{11}{College of Science and Engineering, Kanto Gakuin University, 1-50-1 Mutsuura Higashi, Kanazawa-ku, Yokohama, Kanagawa 236-8501, Japan}
\altaffiltext{12}{Institute for Space-Earth Environmental Research (ISEE), Nagoya University, Furo-cho, Chikusa-ku, Nagoya, Aichi 464-8601, Japan}
\altaffiltext{13}{Department of Astronomy, University of Michigan, Ann Arbor, MI 48109, USA}
\altaffiltext{14}{Department of Physics, Norwegian University of Science and Technology, NO-7491 Trondheim, Norway}
\altaffiltext{15}{Harvard-Smithsonian Center for Astrophysics, 60 Garden Street, Cambridge, MA 02138, USA}


\maketitle

\begin{abstract}
To understand physical processes such as mass transfer and binary evolution in X-ray binaries, the orbital parameters of the system are fundamental and crucial information. Cygnus X-3 is a high-mass X-ray binary composed of a compact object of unknown nature and a Wolf-Rayet star, which is of great interest in the context of wind-fed mass accretion and binary evolution. Here we present XRISM/Resolve high-resolution spectroscopy focusing on the Fe Ly$\alpha$ lines in its hypersoft state. We perform an orbital phase-resolved spectral analysis of the lines to study the orbital modulation of the emission and absorption lines. It is found that the emission lines reflect the orbital motion of the compact object whose estimated velocity amplitude is $430^{+150}_{-140}~\si{km.s^{-1}}$, while the absorption lines show a variation that can be interpreted as originating from the stellar wind. We discuss possible mass ranges for the binary components using the mass function with the estimated value of the velocity amplitude in this work, combined with the relation between the mass loss rate and the orbital period derivative and the empirical mass and mass loss rate relation for Galactic Wolf-Rayet stars. They are constrained to be $(1.3\text{--}5.1)\,M_\odot$ and $(9.3\text{--}12)\,M_\odot$ for the assumed inclination angle of $i =\ang{25}$, which becomes more relaxed to $(1.3\text{--}24)\,M_\odot$ and $(9.3\text{--}16)\,M_\odot$ for $i = \ang{35}$, respectively. Thus, it remains unclear whether the system harbors a black hole or a neutron star.
\end{abstract}


\section{Introduction}
Cygnus X-3 (hereafter Cyg X-3) is a unique Galactic high-mass X-ray binary consisting of a compact object and a WN4-6 type Wolf-Rayet (WR) companion star \citep{vankerkwijk1992,vankerkwijk1993,vankerkwijk1996,koljonen2017}, located $9.7\pm 0.5~\si{kpc}$ from Earth \citep{reid2023}. WR stars have strong stellar winds, making them suitable targets for studying wind accretion. Since there are only two observed Galactic WR X-ray binaries -- Cyg X-3 and OAO 1657-415 \citep{mason2009} -- this system is a valuable target for observational studies of such a system. Indeed, many of the characteristics observed in the X-ray spectra of Cyg X-3 are thought to originate from the strong stellar wind of the WR star (e.g., \cite{kallman2019}). The X-ray spectra display distinct signatures of the photoionized plasma, including emission lines, absorption lines, and narrow radiation recombination continua (RRCs) from highly ionized species of elements such as Si, S, and Fe (e.g., \cite{kawashima1996,paerels2000}). The observation with XRISM, equipped with the microcalorimeter Resolve, has significantly enhanced the amount of information imprinted on the X-ray spectra regarding the photoionized plasma \citep{collaboration2024c}.

WR X-ray binaries are expected to evolve into double compact binaries, which are gravitational wave sources, making their binary interactions a subject of great interest (e.g., \cite{marchant2024a}). The fate of Cyg X-3 was discussed in \citet{belczynski2013}, arguing that the system is a very likely double black hole or black-hole--neutron-star progenitor. In the study of binary evolution and the fundamental understanding of the system itself, orbital parameters represent the most basic and essential information. The orbital period of Cyg X-3 is well known to be $\sim \SI{4.8}{h}$. The orbital inclination of the system has been estimated using several different methods, which consistently yield a value of $\sim \ang{30}$ (\cite{vilhu2009}, photoionized wind modeling; \cite{antokhin2022}, X-ray \& IR photometry; \cite{veledina2024a},\cite{veledina2024}, X-ray polarimetry). \citet{miller-jones2004} also derived an orbital inclination of $\approx \ang{30}$ from the proper motion analysis of the radio jets, whereas the precession modeling suggests a smaller inclination of $\lesssim \ang{14}$. On the other hand, the masses of the compact object and WR star are not well determined, leaving it unclear whether the compact object is a black hole or a neutron star.

A commonly used method for estimating the masses of stars in a binary system is to measure the orbital velocity from the Doppler modulation of emission lines and derive the mass function. There have been several studies measuring the orbital velocity of the compact object in Cyg X-3. \citet{vilhu2009} analyzed Chandra/HETG observational data and investigated the Doppler modulation of Fe Ly$\alpha$ emission lines. The Ly$\alpha$ emission in the photoionized plasma predominantly originates from cascade processes after the recombination between bare Fe nuclei and free electrons, which takes place in the regions closest to the compact object. Therefore, these lines serve as a probe of the motion of the compact object (although the degree of ionization also depends on the electron density). Based on this idea, \citet{vilhu2009} measured the velocity amplitude of its circular motion to be $418\pm \SI{123}{km.s^{-1}}$. However, this measurement is based on a modeling that does not account for the absorption lines detected by the Resolve \citep{collaboration2024c}, as they could not be resolved with the spectral resolution of HETG. On the other hand, \citet{collaboration2024c} conducted an orbital phase-resolved spectral analysis of the Fe K band using photoionized plasma modeling and estimated the orbital velocity of the compact object to be $194\pm \SI{29}{km.s^{-1}}$ based on the Doppler shift of the emission line components. It is worth noting that the velocity amplitude of $\sim \SI{200}{km.s^{-1}}$ strongly supports the presence of a black hole in this system, as a lower velocity is expected for a more massive compact object. In that analysis, however, an assumption was made that all emission line components with different ionization states share a common velocity, and thus the velocity primarily determined by Fe He$\alpha$, not Fe Ly$\alpha$, because of the higher photon statistics in the former. Since a more highly ionized plasma is expected to be formed closer to and probably confined by the compact object, it is ideal to use the energy shift of the Fe Ly$\alpha$ lines for the velocity measurement.

In this paper, we focus exclusively on the Fe Ly$\alpha$ spectra observed with XRISM/Resolve and perform the measurement of the radial velocity of the compact object in Cyg X-3 by using a model that accounts for the presence of absorption lines. Details of the observation and data reduction are described in Section \ref{sec:obs_data}. The analysis and results are presented in Section \ref{sec:anaysis_results}, followed by a discussion of their implications in Section \ref{sec:discussion}. Finally, we conclude this study in Section \ref{sec:conclusion}.

\section{Observation and Data Reduction}\label{sec:obs_data}
The XRISM observation of Cyg X-3 was conducted as one of the Performance Verification (PV) observations, starting at 06:53:50 UT on 2024 March 24 until 15:37:46 UT on 2024 March 25 when the source was in a "hypersoft" state (ObsID: 300065010). To reduce the data, we followed the procedures described in the XRISM Quick-Start Guide ver.\,2.3 using tools in HEAsoft ver.\,6.34 and the calibration database (CALDB) ver.\,8 (v20240815). The effective exposure after the screening was $\approx \SI{67}{ks}$. 
The spectrum was extracted from High primary (Hp) grade events from all pixels except pixel 27 due to known calibration uncertainties. Large (L-type) Redistribution Matrix Files (RMFs) are created with the {\tt rslmkrmf} task using the cleaned event file. Auxiliary Response Files (ARFs) are generated with the {\tt xaarfgen} task.

\section{Analysis and Results}\label{sec:anaysis_results}
To perform phase-resolved spectroscopy, we divide the orbital period into eight phase bins with equal width (i.e., $0.125$ for each bin) by adopting the quadratic ephemeris of \citet{antokhin2019} with an orbital period of $\SI{17252}{s}$ and extract spectra from each bin. Certain phases are always subject to visibility constraints due to the orbital period of Cyg X-3 being a multiple of the period over which XRISM orbits the Earth, and therefore phase $\varphi = 0.250\text{--}0.375$ is severely underexposed and unsuitable for spectral analysis. 

\begin{figure*}[htbp]
  \begin{tabular}{cccc}
    \begin{minipage}{0.25\textwidth}
      \centering
      \includegraphics[width=\linewidth]{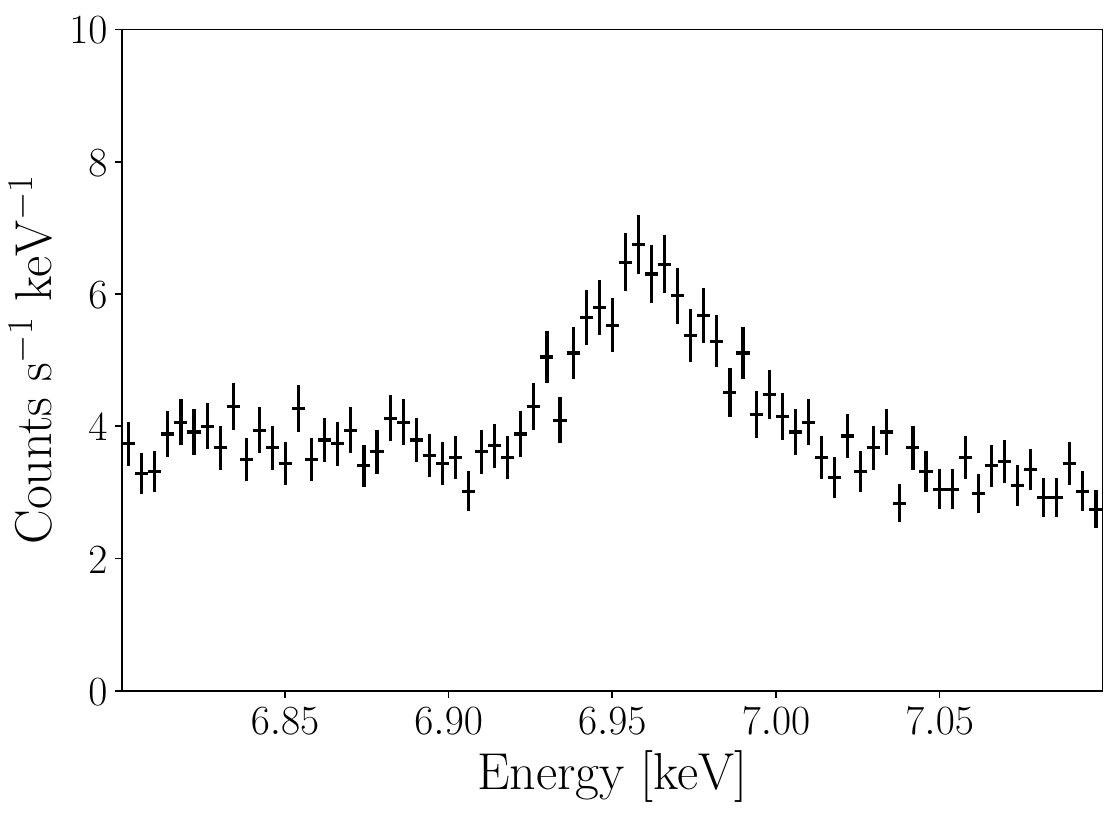}
      \subcaption{$\varphi = 0.000\text{--}0.125$}
    \end{minipage}
    \begin{minipage}{0.25\textwidth}
      \centering
      \includegraphics[width=\linewidth]{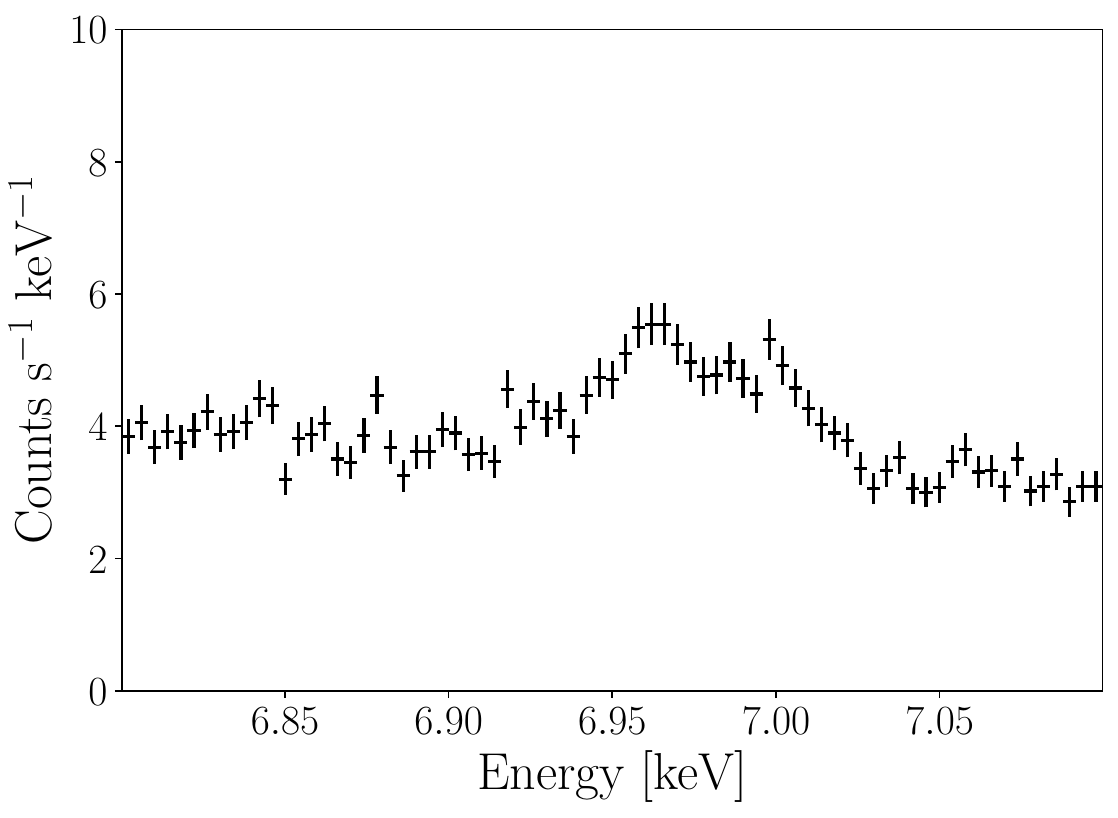}
      \subcaption{$\varphi = 0.125\text{--}0.250$}
    \end{minipage}
    \begin{minipage}{0.25\textwidth}
      \centering
      \includegraphics[width=\linewidth]{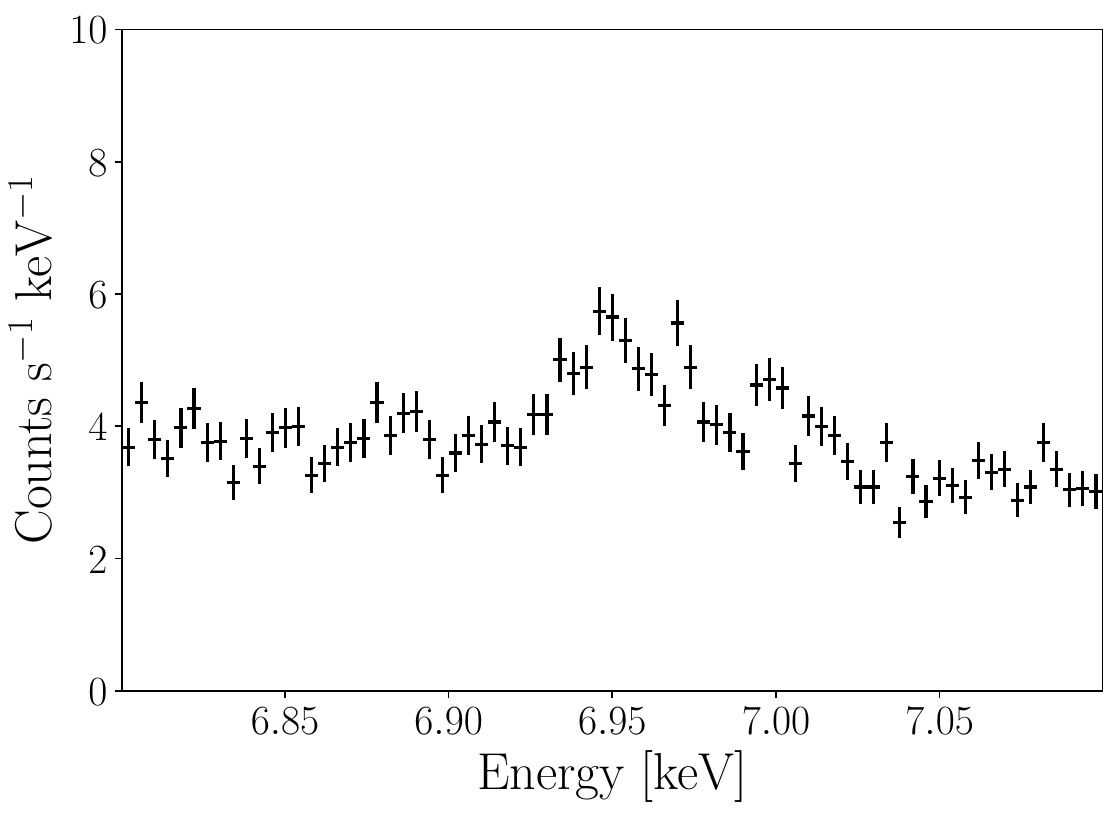}
      \subcaption{$\varphi = 0.375\text{--}0.500$}
    \end{minipage}
    \begin{minipage}{0.25\textwidth}
      \centering
      \includegraphics[width=\linewidth]{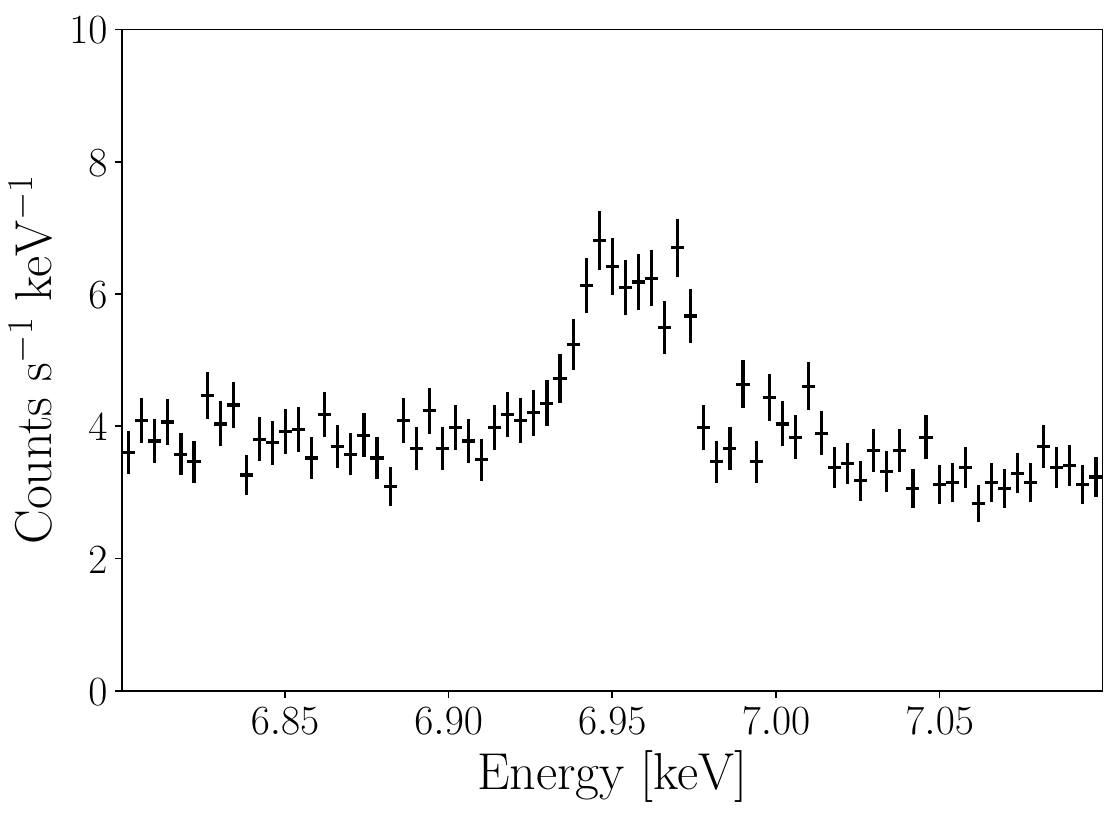}
      \subcaption{$\varphi = 0.500\text{--}0.625$}
    \end{minipage}\\
    \begin{minipage}{0.25\linewidth}
      \centering
      \includegraphics[width=\linewidth]{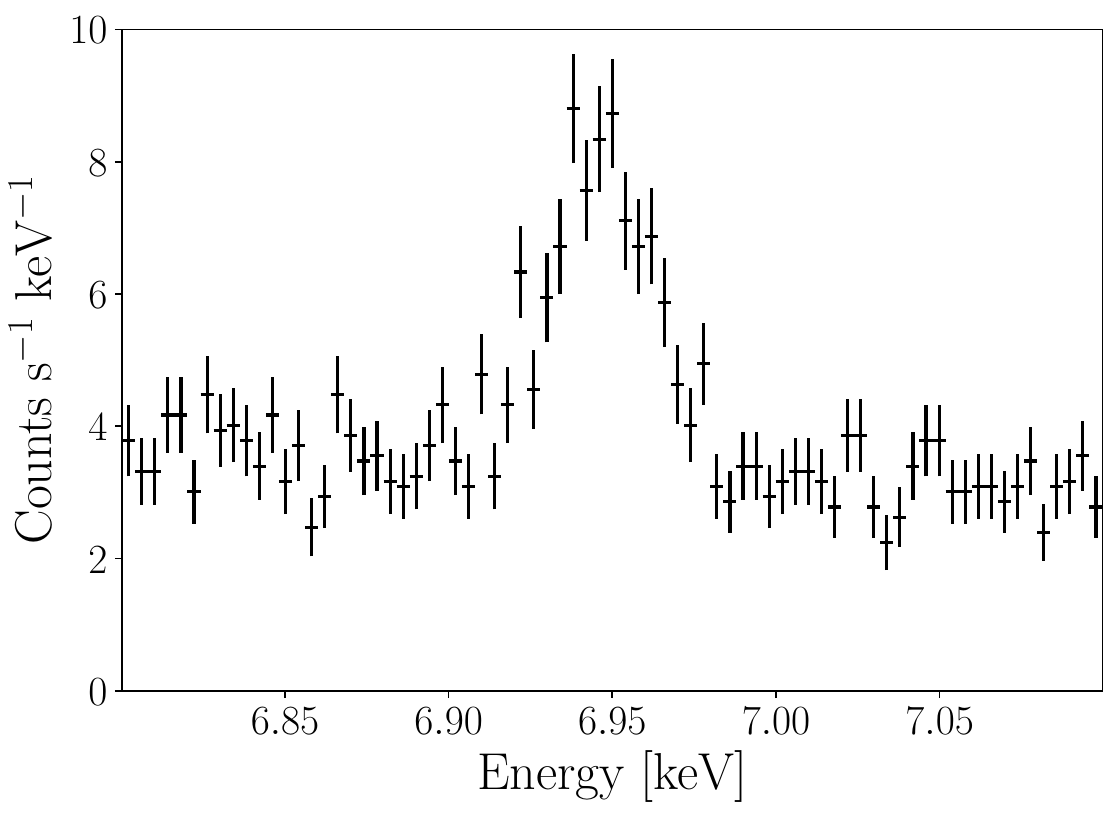}
      \subcaption{$\varphi = 0.625\text{--}0.750$}
    \end{minipage}
    \begin{minipage}{0.25\linewidth}
      \centering
      \includegraphics[width=\linewidth]{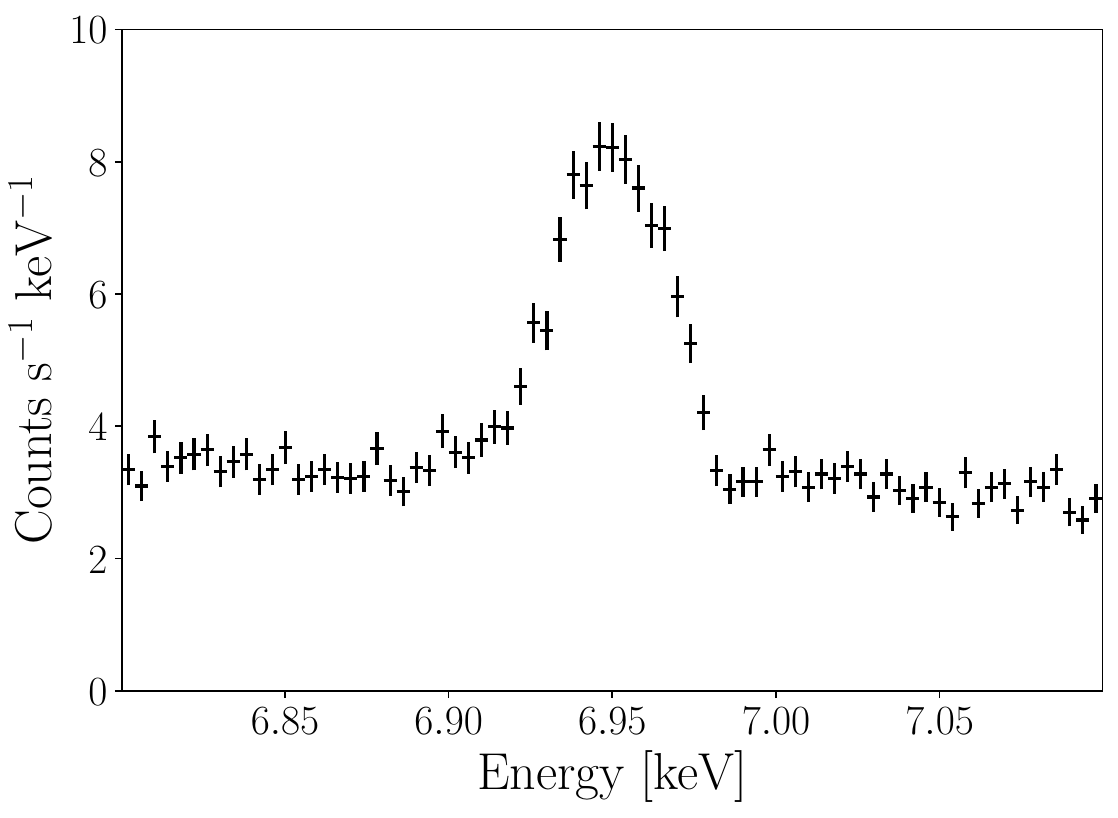}
      \subcaption{$\varphi = 0.750\text{--}0.875$}
    \end{minipage}
    \begin{minipage}{0.25\linewidth}
      \centering
      \includegraphics[width=\linewidth]{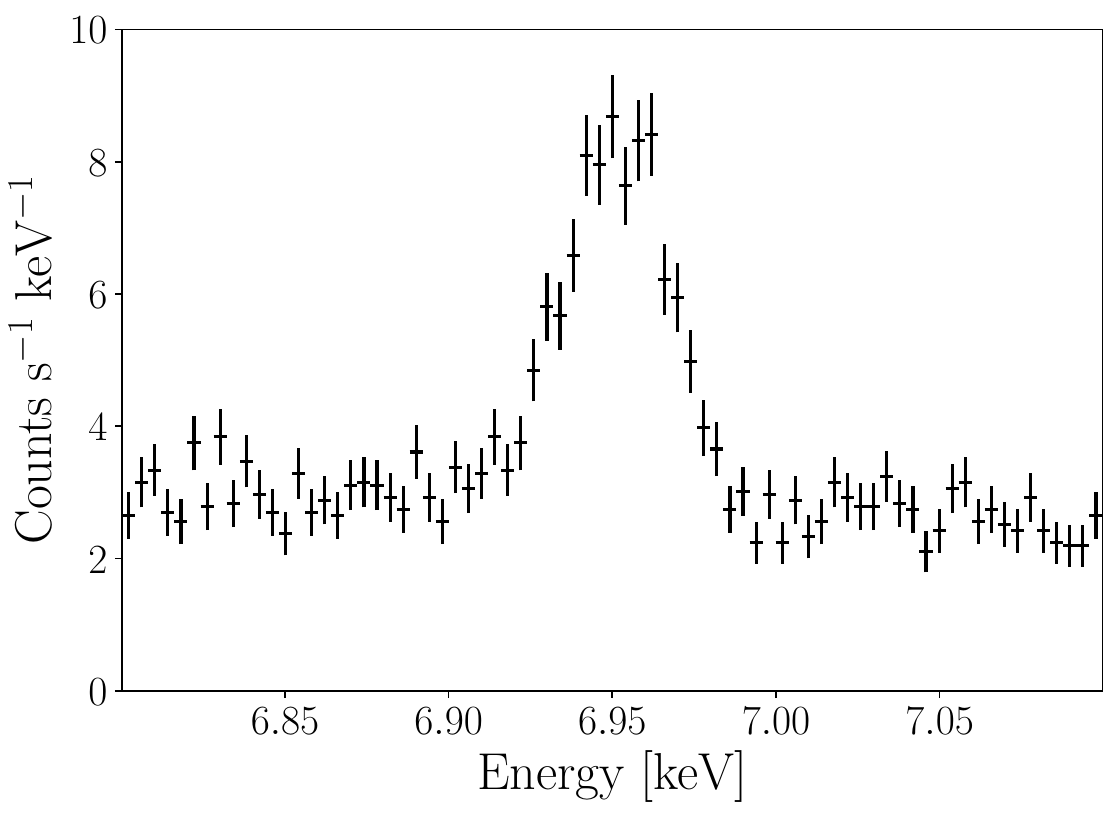}
      \subcaption{$\varphi = 0.875\text{--}1.000$}
    \end{minipage}
  \end{tabular}
  \caption{
    Orbital phase-resolved spectra of Fe Ly$\alpha$ lines ($6.8\text{--}\SI{7.1}{keV}$). The spectrum extracted from orbital phase $\varphi = 0.250\text{--}0.375$ is removed because of severe underexposure due to the visibility constraint. The spectra are rebinned into $\SI{4}{eV}$ bins for clarity, although the unbinned spectra are used for the analysis.
  }
  \label{fig:spectra}
\end{figure*}

The Fe Ly$\alpha$ spectra from the seven different phases are shown in Figure \ref{fig:spectra}. These spectra show that the Fe Ly$\alpha$ profile is formed by the emission and absorption lines overlapping with each other. This overlap leads to parameter distributions with some local extrema, making it difficult to find the optimal solution using conventional optimization methods such as maximum likelihood estimation with C-statistics. Therefore, due to its flexibility, we adopt Bayesian inference based on the Markov Chain Monte Carlo (MCMC) method in the subsequent analysis.

\subsection{Analyzing each phase independently}\label{sec:analysis_ind}
As a first step, we analyze the $6.8\text{--}\SI{7.1}{keV}$ spectrum extracted from each phase independently. The continuum component is modeled with a power law whose free parameters are the normalization $K$ ($\si{photons.s^{-1}.cm^{-2}.keV^{-1}}$ at $\SI{1}{keV}$) and the photon index $\alpha$. The emission and absorption lines are modeled with four Gaussian functions. Two of them correspond to the Ly$\alpha_1$ and Ly$\alpha_2$ emission, and the other two correspond to the Ly$\alpha_1$ and Ly$\alpha_2$ absorption. We fix the Ly$\alpha_1$ and Ly$\alpha_2$ line energies to their rest-frame values ($\SI{6.973}{keV}$ and $\SI{6.952}{keV}$, respectively) and set redshift $z$ as free parameters. We also fix the Ly$\alpha_1 /$Ly$\alpha_2$ flux ratios of both emission and absorption lines to 2. Removing this assumption does not significantly affect the subsequent results or discussion. The redshift and width of the Ly$\alpha_1$ and Ly$\alpha_2$ lines are assumed to be identical within the emission or absorption component, but are treated independently between the emission and absorption. In summary, the free parameters of emission and absorption components are the redshift of emission and absorption lines ($z_{\rm e}$ and $z_{\rm a}$), line width ($\sigma_{\rm e}$ and $\sigma_{\rm a}$) and the normalization of the Ly$\alpha_2$ line ($N_{\rm e}$ and $N_{\rm a}$).

To optimize prior distributions for the parameters given above, we conduct an ad hoc spectral fitting based on $C$-statistic \citep{cash1979} in XSPEC \citep{arnaud1996}. Only in this step, we fit the seven spectra simultaneously, assuming the common power-law normalization $K$ and photon index $\alpha$ across the phases (i.e., in the subsequent MCMC analysis, these parameters are independent among the phases). The prior distributions optimized with this approach are summarized in Table \ref{tab:prior}. The parameters of the continuum component are sampled from a normal distribution with the mean values from this optimization and a standard deviation of $1$. The line parameters are sampled from a uniform distribution over a range such that all values obtained by the optimization for each phase are included. 


\begin{table}
  \caption{Prior distributions of the model parameters.}
  \label{tab:prior}
  \centering
  \begin{threeparttable}
    \begin{tabularx}{\linewidth}{>{\raggedright\arraybackslash}X >{\centering\arraybackslash}X}\hline\hline\noalign{\vskip3pt}
      Parameter & Prior distribution  \\ \hline\noalign{\vskip3pt}
      \multicolumn{2}{c}{Continuum} \\ \hline\noalign{\vskip3pt}
      Norm ($K$)& $\mathcal{N}(10.3,1)$\tnote{*} \\
      Photon index ($\alpha$)& $\mathcal{N}(2.75,1)$ \\ \hline\noalign{\vskip3pt}
      \multicolumn{2}{c}{Emission lines} \\ \hline\noalign{\vskip3pt}
      Norm ($N_{\rm e}$) & $\mathcal{U}(0,\num{5e-3})$\tnote{**} \\
      Width ($\sigma_{\rm e}$) [$\si{keV}$] & $\mathcal{U}(\num{5e-3},\num{3e-2})$ \\
      Redshift ($z_{\rm e}$) & $\mathcal{U}(\num{-2e-3},\num{3e-3})$ \\ \hline\noalign{\vskip3pt}
      \multicolumn{2}{c}{Absorption lines} \\ \hline\noalign{\vskip3pt}
      Norm ($N_{\rm a}$)& $\mathcal{U}(\num{-2e-3},0)$ \\
      Width ($\sigma_{\rm a}$) [$\si{keV}$] & $\mathcal{U}(\num{1e-3},\num{3e-2})$ \\
      Redshift ($z_{\rm a}$) & $\mathcal{U}(\num{-2e-3},\num{0})$ \\ \hline
    \end{tabularx}
    \begin{tablenotes}
      \item[*] $\mathcal{N}(\mu,\sigma)$ represents a normal distribution with a mean of $\mu$ and a standard deviation of $\sigma$.
      \item[**] $\mathcal{U}(x_l,x_h)$ represents a uniform distribution in the range from $x_l$ to $x_h$.
    \end{tablenotes}
  \end{threeparttable}
\end{table}

We then perform a Hamiltonian Monte Carlo (HMC; e.g., \cite{neal2011}) No-U-Turn Sampler (NUTS; \cite{JMLR:v15:hoffman14a}) implemented in NumPyro \citep{phan2019a}, with 1000 warm-up steps for the adoption phase of HMC and 2000 samples. We set the maximum number of leapfrog steps at each iteration to be $1023$. We use the Gelman-Rubin diagnostic \citep{gelman1992} to check convergence, and the condition $\hat{R} \leq 1.05$ for all parameters indicates that the MCMC chains have sufficiently converged. 

\begin{figure*}
  \begin{center}
      \includegraphics[width=0.7\linewidth]{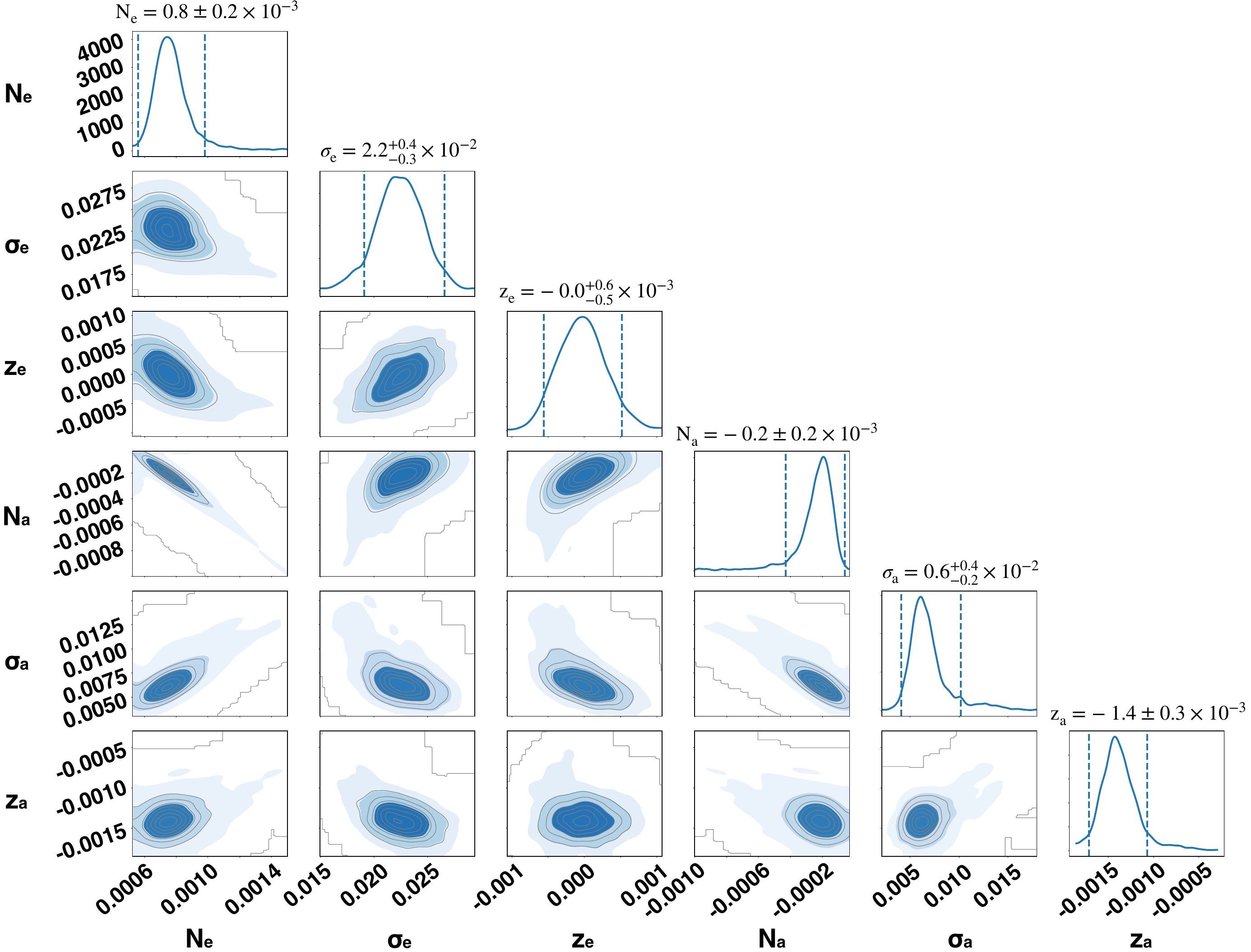}
  \end{center}
  \caption{Corner plot of the posterior samples from $\varphi=0.375\text{--}0.500$ when analyzing each phase independently(\S\ref{sec:analysis_ind}). The vertical dashed lines indicate the 90\% HPDIs.}
  \label{fig:corner_ind_bin4}
\end{figure*}

The prediction (median) with the $90\%$ highest probability density interval (HPDI) from the fitting for $\varphi=0.375\text{--}0.500$ overlaid on the observed spectrum is shown in Figure \ref{fig:spec_ind_bin4}. In Figure \ref{fig:spec_ind} in appendix \ref{sec:appendix_additional}, we also show those for all seven phases. The corner plot of posterior samples for $\varphi = 0.375\text{--}0.500$ is shown in Figure \ref{fig:corner_ind_bin4}. There are some clear positive and negative correlations between the parameters. For example, as can be easily expected, the normalizations of emission and absorption lines ($N_{\rm e}$ and $N_{\rm a}$) are anticorrelated. 
The redshift of the emission lines $z_{\rm e}$, which is of the most interest in this study, correlates with $N_{\rm a}$ and $\sigma_{\rm e}$, but anticorrelates with $N_{\rm e}$ and $\sigma_{\rm a}$.

\begin{figure}
    \begin{center}
        \includegraphics[width=\linewidth]{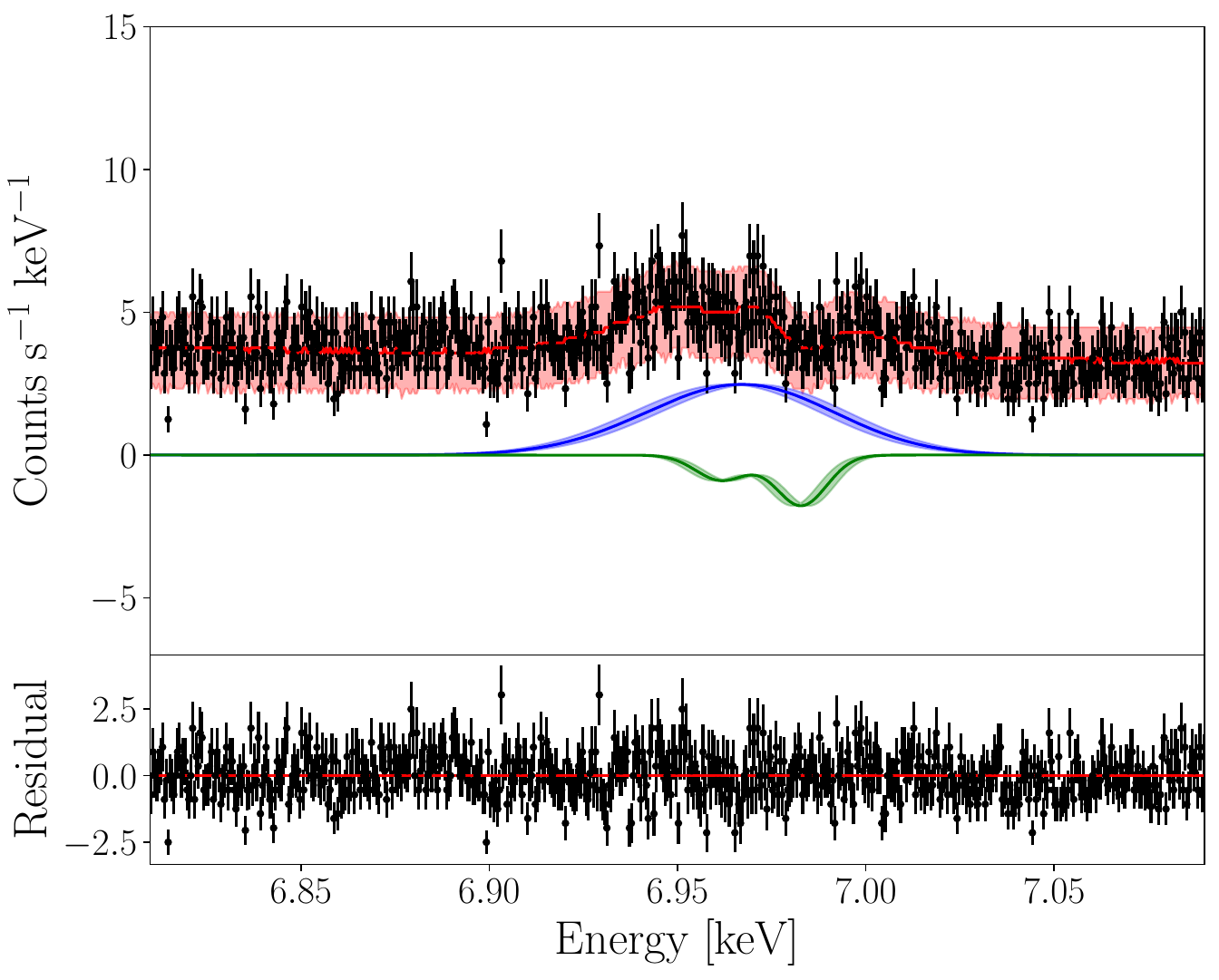}
    \end{center}
    \caption{Top: The prediction (median) with the $90\%$ HPDI from the fitting for $\varphi=0.375\text{--}0.500$ overlaid on the observed spectrum. The red solid line represents the predicted median spectrum and red-shaded region represents $90\%$ HPDI. The blue and green solid lines show the emission and absorption lines with the medians of posterior samples of the parameters described in Table \ref{tab:median_hpdi_ind} in appendix \ref{sec:appendix_additional}. The blue- and green-shaded region correspond to the $90\%$ HPDIs for redshifts ($z_{\rm e}$ and $z_{\rm a}$). Bottom: The residual from the median.}
    \label{fig:spec_ind_bin4}
\end{figure}

\begin{figure}
    \begin{center}
        \includegraphics[width=\linewidth]{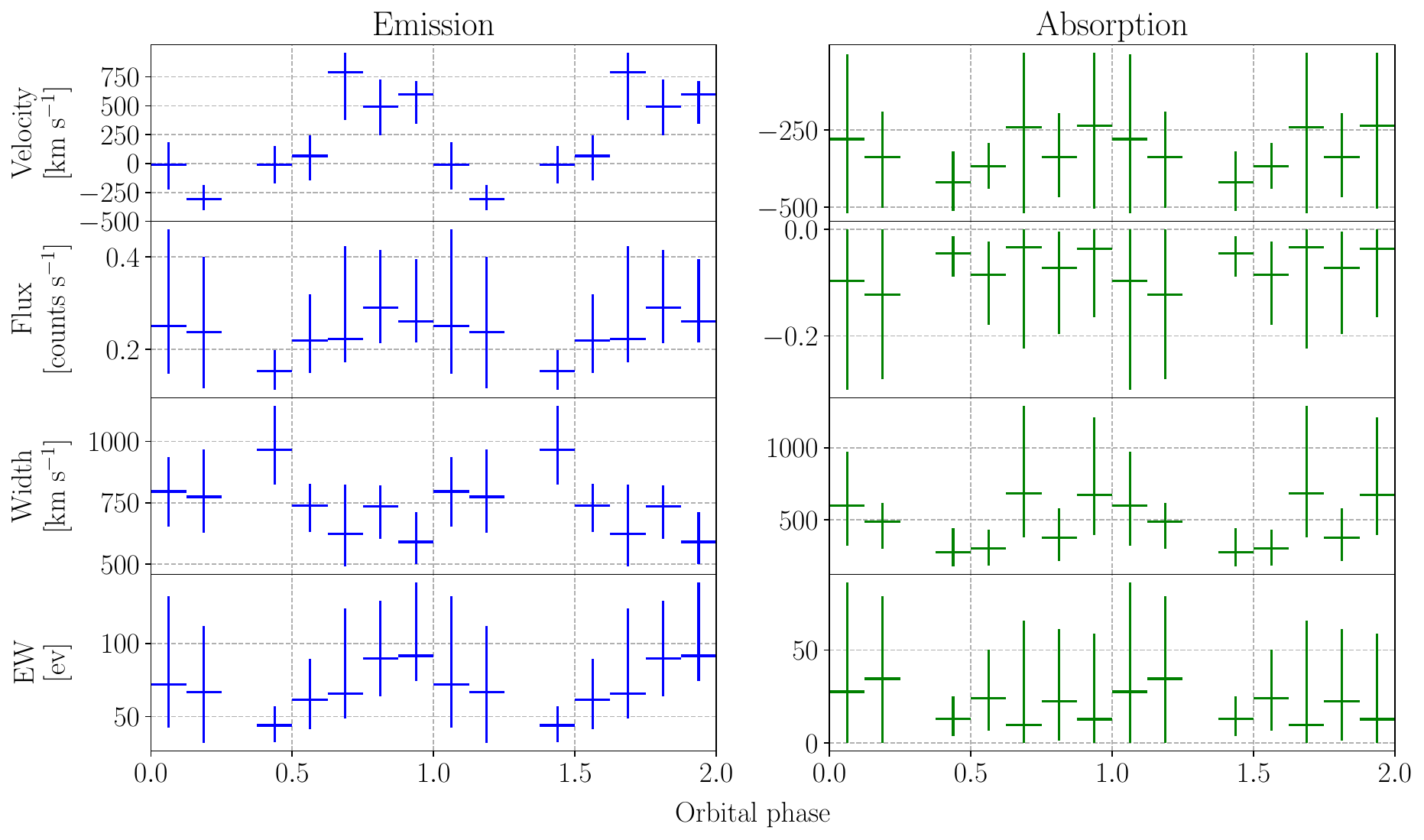}
    \end{center}
    \caption{Line parameters obtained for the emission (left) and absorption (right) components as a function of the orbital phase when analyzing each phase independently (\S\ref{sec:analysis_ind}). From top to bottom, they represent the velocities, fluxes, widths, and equivalent widths (EWs).}
    \label{fig:param_phase_dep_ind}
\end{figure}

Table \ref{tab:median_hpdi_ind} in appendix \ref{sec:appendix_additional} provides the medians of the marginal posteriors and their $90\%$ HPDIs. This table also presents the $C$-statistics with corresponding degrees of freedom (dof), calculated using the medians of the model predictions obtained through the HMC method, for the purpose of evaluating the model quality. We heve obtained acceptable $C$-statistic/dof values. Figure \ref{fig:param_phase_dep_ind} shows the resulting velocity, flux, width, and equivalent width (EW) as a function of orbital phase for each of the emission and absorption lines, where $v=cz$ ($c$ is the speed of light) is used to convert redshift to velocity. The velocity of the emission line component shows a significant orbital modulation, with the largest blueshift at around phase $0.25$ and the largest redshift at around $0.75$, indicating that the Fe Ly$\alpha$ emission lines originate from the vicinity of the compact star, and the velocity shifts track the orbital motion of the compact star as expected.

\subsection{Analyzing all the phases simultaneously assuming a circular orbit} \label{sec:analysis_sine}
As demonstrated in the previous section, the velocity shift of the emission lines indicates the orbital motion of the compact star. 
We therefore introduce more sophisticated approach, where the spectra of all the phases are analyzed simultaneously, assuming a circular motion for the velocity shift of the emission line component. Specifically, the redshift of the Gaussian is represented by the following equation:
\begin{equation}
    z_{\rm e} = \frac{v_0}{c} - \frac{v}{c}\sin\qty{2\pi(\varphi+\varphi_0)} \label{sine}
\end{equation}
where $v_0,~v,~\varphi$ and $\varphi_0$ are the velocity offset, the velocity amplitude, the orbital phase, and the phase offset, respectively. \citet{antokhin2019} investigated the orbital period and its variation using observational data from multiple X-ray telescopes and detected a sinusoidal perturbation in the period. If this perturbation is interpreted as apsidal motion, the eccentricity of the orbit is estimated to be $\approx 0.03$. Therefore, assuming a circular orbit should be reasonable. 

Based on the assumption outlined above, the analysis is performed in the same manner as in the previous section. The priors for the velocity offset, the velocity amplitude, and the phase offset are sampled from a uniform distribution with the range of $[-100,~300]~\si{km.s^{-1}},~[200,~700]~\si{km.s^{-1}},~[-0.15,~0.15]$, respectively. For the other parameters, the priors are sampled using the same method as in the previous section. The orbital phase $\varphi$ is fixed at the median of the phase range from which each spectrum is extracted for each bin. We perform an HMC-NUTS with 1000 warm-up steps and 2000 samples. All parameters satisfy the condition $\hat{R} \leq 1.05$, which indicates sufficient convergence of the MCMC chains.

\begin{figure*}
  \begin{center}
      \includegraphics[width=0.7\linewidth]{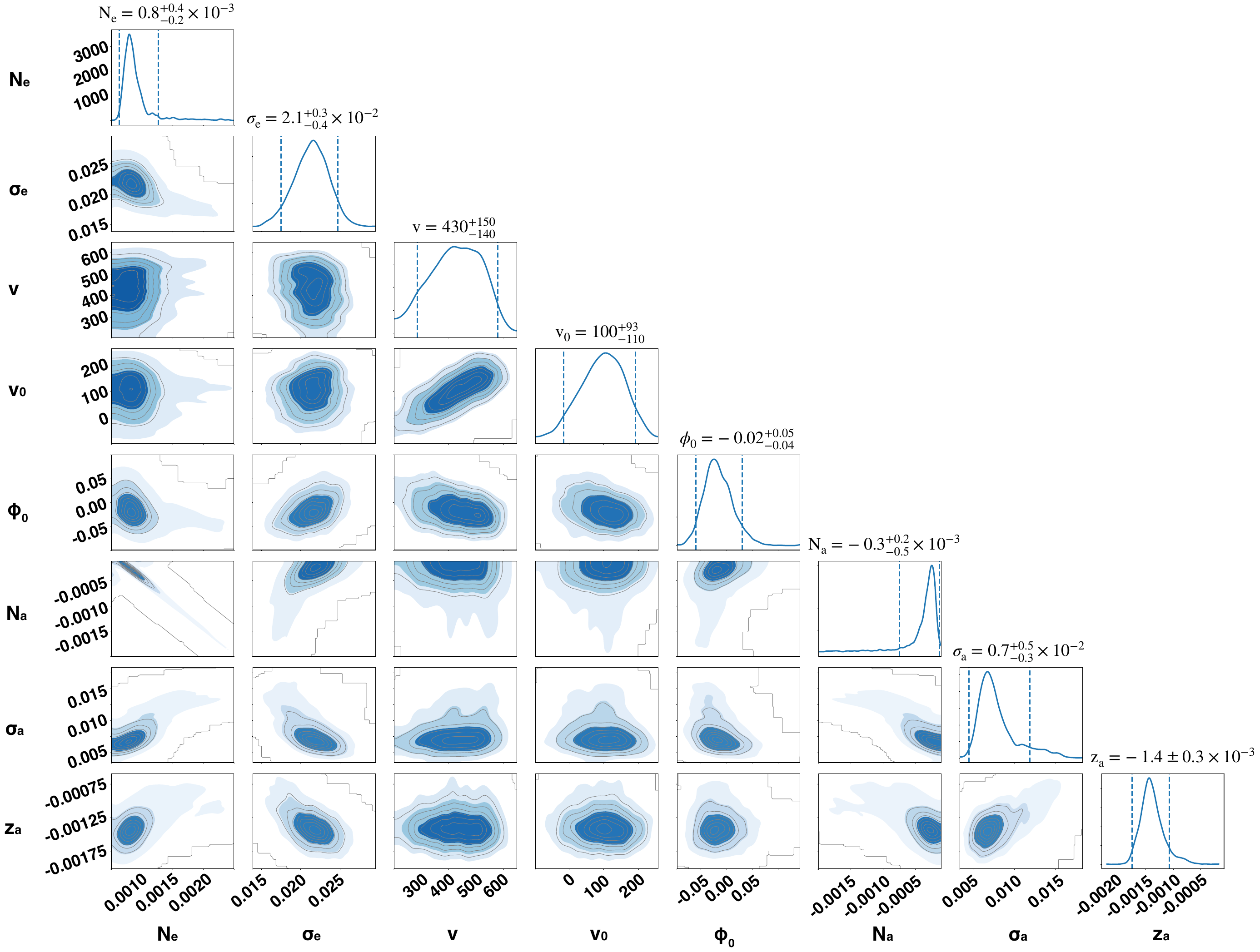}
  \end{center}
  \caption{The same figure as Figure \ref{fig:corner_ind_bin4} but obtained by the analysis assuming a circular orbit(\S\ref{sec:analysis_sine}).}
  \label{fig:corner_sine_bin4}
\end{figure*}

\begin{figure}[h]
    \begin{center}
        \includegraphics[width=\linewidth]{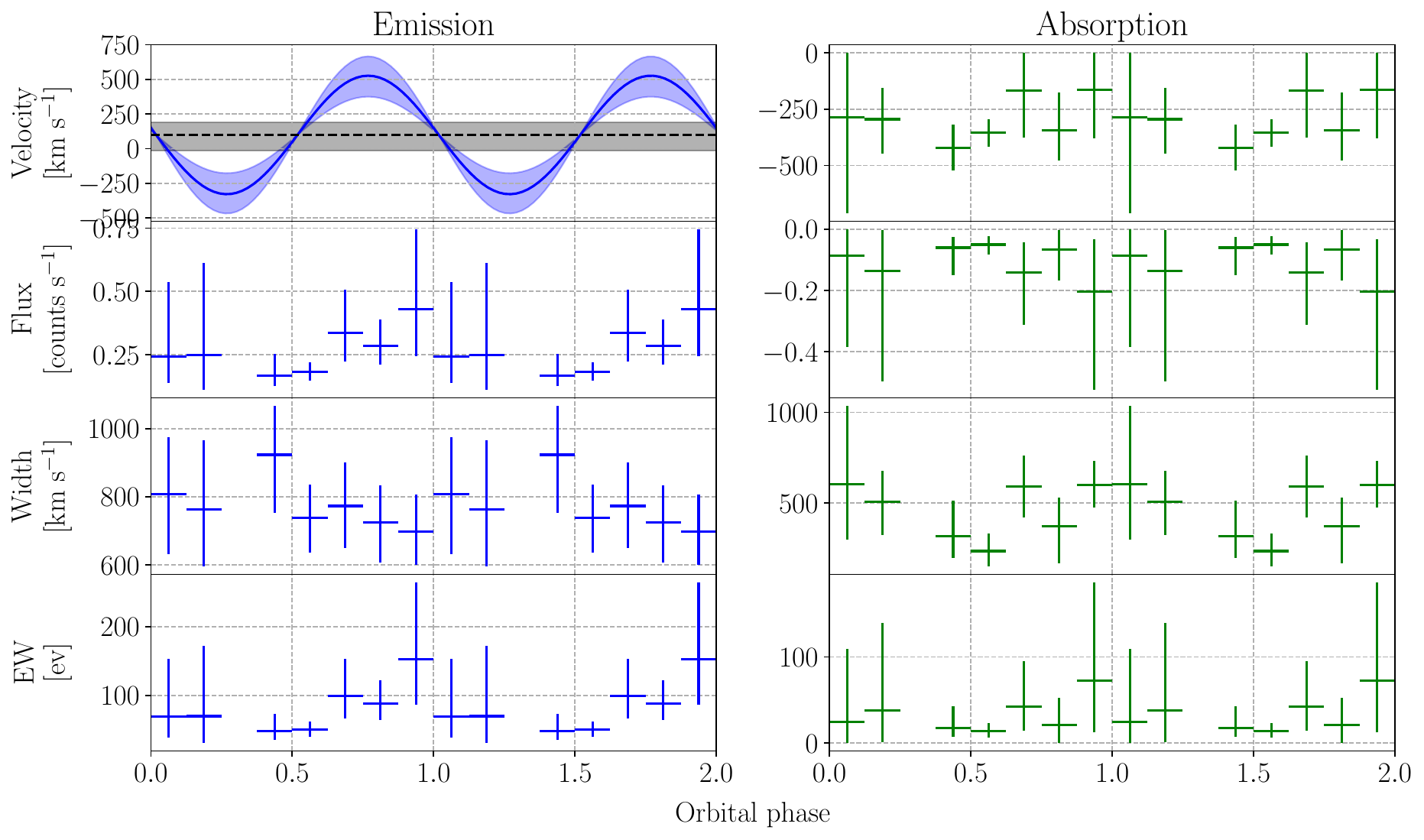}
    \end{center}
    \caption{The same figure as Figure \ref{fig:param_phase_dep_ind} except for the velocity of the emission component (top-left panel), but obtained by the analysis assuming a circular orbit. The blue solid line represents the velocity curve of the emission component with the median of the posterior samples of the velocity amplitude, the velocity offset, and the phase offset. The blue-shaded region corresponds to the $90\%$ HPDI of the velocity amplitude. The gray dashed line and the gray-shaded region represent the median and the $90\%$ HPDI of the posterior samples of the velocity offset, respectively.}
    \label{fig:param_phase_dep_sine}
\end{figure}

The predictions (median) with the $90\%$ HPDIs from the fitting for all phases overlaid on the observed spectra are shown in Figure \ref{fig:spec_sine} in appendix \ref{sec:appendix_additional}. The corner plot of the posterior samples is also shown in Figure \ref{fig:corner_sine_bin4}. Figure \ref{fig:param_phase_dep_sine} shows the resulting parameters as a function of the orbital phase, while Table \ref{tab:median_hpdi_sine} in appendix \ref{sec:appendix_additional} provides the medians of the marginal posteriors and their $90\%$ HPDIs with the $C$-statistic and dof. We have obtained an acceptable value of $C$-statistic/dof value. The velocity amplitude, the velocity offset, and the phase offset are $v = 430^{+150}_{-140}~\si{km.s^{-1}}$, $v_0 = 100^{+93}_{-110}~\si{km.s^{-1}}$, and $\varphi_0 = -0.02^{+0.05}_{-0.04}$, respectively. The flux of the emission line seems to be minimum in orbital phase $0.5$ and maximum in $0.0$. The width of the emission line remains roughly constant, though it is slightly larger in $\varphi=0.385\text{--}0.500$. The absorption lines are overall blueshifted by $> \SI{250}{km.s^{-1}}$ and appear to show a small-amplitude modulation, with the blueshift being minimal around $\varphi=0.0$ and maximal around $\varphi=0.5$. The flux variation is not clear.

\section{Discussion}\label{sec:discussion}
\subsection{Summary of the Results and Comparison with Previous Work}\label{sec:result_summary}
We have performed an orbital phase-resolved analysis of the X-ray spectra of Cyg X-3 in its hypersoft state obtained by XRISM. To measure the radial velocity of the compact star, we have conducted the analysis focusing on the Fe Ly$\alpha$ emission and absorption lines. By analyzing the spectra extracted from each phase independently, we have confirmed that, as expected, the velocity shift of the emission lines reflects the orbital motion of the compact object. We have proceeded with the analysis by assuming the velocity shift of the emission lines follows a sine curve, as described in Equation \ref{sine}. As a result, the orbital velocity amplitude of the compact object is determined to be $v = 430^{+150}_{-140}~\si{km.s^{-1}}$. The velocity offset includes $\SI{0}{km.s^{-1}}$ within the $90\%$ HPDI. These estimated values are consistent with those reported in \citet{vilhu2009}, though the absorption lines were not taken into account in their analysis due to the insufficient spectral resolution of Chandra/HETG. 

\citet{collaboration2024c} analyzed the same dataset as this study. As mentioned earlier, their analysis of the Fe K band spectra employed the model with three photoionized plasma components of different ionization degrees for both emission and absorption lines. The three emission components and the three absorption components were each constrained to share a common Doppler shift. As a result, the orbital velocity amplitude of the emission components was found to be $194\pm \SI{29}{km.s^{-1}}$, significantly smaller than the values obtained in this study and by \citet{vilhu2009}. This was likely because the velocity was primarily determined by the less ionized component, which represents the complex spectral structure of the Fe He$\alpha$ emission. Our result suggests that the most highly ionized plasma emitting Fe Ly$\alpha$ lines is spatially separated from the less ionized plasma responsible for the Fe He$\alpha$ emission and exhibits distinct motion.

The same applies to the absorption lines, whose Doppler shift is primarily determined by the low-ionization component in \citet{collaboration2024c}. In the previous study, the velocity amplitude and offset of the absorption lines are reported as $55\pm\SI{7}{km.s^{-1}}$ and $-534\pm \SI{6}{km.s^{-1}}$, respectively. The phase at which the blueshift is maximum is $\varphi\sim0.1$. In this study, however, the systemic velocity of the absorption line is found to be $\sim \SI{-300}{km.s^{-1}}$, whose absolute value is smaller than the result from \citet{collaboration2024c}. The modulation pattern also differs, with the phase at which the maximum blueshift occurs being $\varphi \sim 0.5$ (Figure \ref{fig:param_phase_dep_sine}). 

\begin{figure}
  \begin{center}
    \includegraphics[width=\linewidth]{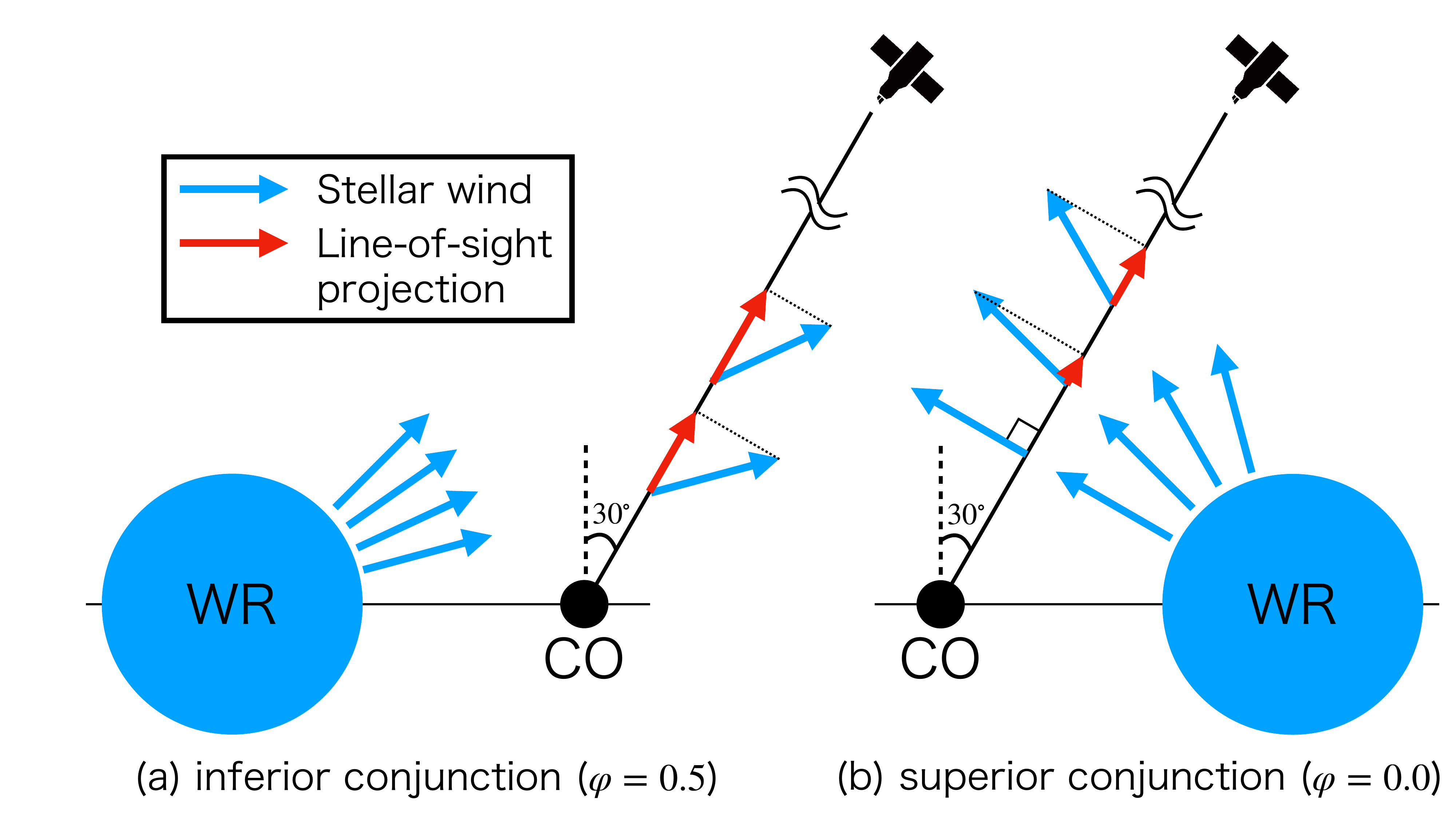}
  \end{center}
  \caption{
    Schematic illustration of the possible stellar wind structure of the WR star in Cyg X-3. (a) corresponds to the inferior conjunction ($\varphi=0.5$), and (b) to the superior conjunction ($\varphi=0.0$) configuration. The blue arrows represent the stellar wind velocity vectors, and the red arrows represent their line-of-sight projections.}
  \label{fig:abs_structure}
\end{figure}

The absorption line width shows a slight orbital modulation with the maximum and minimum at $\varphi\sim0.0$ and $\varphi\sim0.5$, respectively, while the absorption line flux does not show a clear modulation (Figure \ref{fig:param_phase_dep_sine}). Based on the orbital variations of the absorption line parameters described above, it can be interpreted that the absorption lines are formed in the stellar wind of the donor WR star (Figure \ref{fig:abs_structure}). At the inferior conjunction ($\varphi = 0.5$), the absolute value of the stellar wind line-of-sight velocity is maximized. At the superior conjunction ($\varphi = 0.0$), the observed line width is maximum because of the velocity gradient along the line of sight.
The smaller blueshift derived from the Ly$\alpha$ absorption lines in this work can also be naturally explained by the configuration illustrated in Figure \ref{fig:abs_structure}; the higher the ionization (i.e., the closer to the compact object), the smaller the projected component along the line of sight. A careful investigation of various other absorption lines is an important future task for revealing the detailed velocity structure of the stellar wind.

\subsection{Constraints on the System Masses}\label{subsec:mass_constraint}

\begin{figure*}
  \begin{center}
      \includegraphics[width=\linewidth]{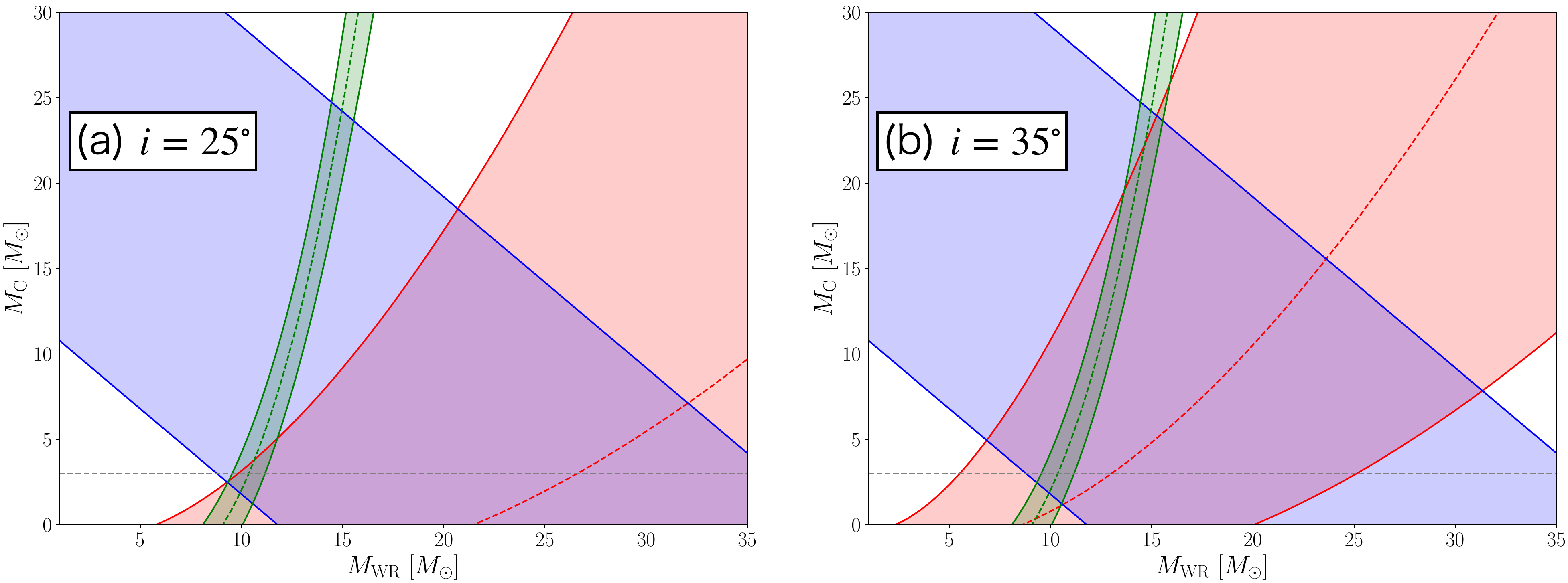}
  \end{center}
  \caption{
    The allowed masses of the compact object and the WR star. The red regions are obtained from Equation \ref{eq:mass_WR} with the orbital velocity amplitude of the compact object obtained in this study, where panel (a) assumes an inclination angle of $i=\ang{25}$ and panel (b) assumes $i=\ang{35}$. The blue and green regions are obtained from Equations \ref{eq:mass_loss_orbital_period_derivative} and \ref{eq:mass_loss_orbital_period_empirical_WR}, respectively. The horizontal dashed lines represent $3\,M_\odot$, theoretically predicted as the upper mass limit of neutron stars.
  }
  \label{fig:mass_constraint}
\end{figure*}

The orbital velocity amplitude of the compact object is related to the mass of the compact object $M_{\rm C}$ and the mass of the WR star $M_{\rm WR}$ through the mass function
\begin{equation}
  \frac{P_{\rm orb}K_{\rm C}^3}{2\pi G} = \frac{M_{\rm WR}^3\sin^3 i}{M_{{\rm tot}}^2} \label{eq:mass_WR}
\end{equation}
where $P_{\rm orb}=\SI{17252}{s}$ is the orbital period, $K_{\rm C}$ is the velocity amplitude of the compact object, $G$ is the gravitational constant, $i$ is the inclination angle of the system, $M_{\rm C}$ is the mass of the compact object, $M_{\rm WR}$ is the mass of the WR star, and $M_{\rm tot} = M_{\rm C} + M_{\rm WR}$ is the total mass of the system. In the following discussion, we assume $i = \ang{25}\text{--}\ang{35}$(\cite{vilhu2009,antokhin2022,veledina2024a,veledina2024}). 
In general, measurement of the velocity amplitude of the WR star $K_{\rm WR}$ with optical or infrared spectroscopy determines the masses of both compact object and WR star using another mass function
\begin{equation}
  \frac{P_{\rm orb}K_{\rm WR}^3}{2\pi G} = \frac{M_{\rm C}^3\sin^3 i}{M_{{\rm tot}}^2} \label{eq:mass_C}
\end{equation}
along with Equation \ref{eq:mass_WR}. However, this method cannot be applied to this system. The position of Cyg X-3 in the Galactic plane leads to strong interstellar extinction, making it impossible to detect its optical/UV counterpart. Therefore, we cannot measure the velocity amplitude of the WR star with optical or UV spectroscopy. Infrared spectroscopy is available. \citet{hanson2000} indeed detected the He{\sc i} 2p-2s $\SI{2.058}{\micro\meter}$ absorption line using the FSpec infrared spectrometer at Steward Observatory, which was then used to measure the velocity amplitude of the WR star. However, with Gemini/GNIRS infrared spectroscopy, \citet{koljonen2017} concluded that the absorption line is not due to the WR star but is formed by the stellar wind. This calls into question the reliability of the published mass function of Cyg X-3 (e.g., \cite{zdziarski2013}). Furthermore, they argued that the infrared emission lines are from a region larger than the binary orbit, making it difficult to determine the orbital parameters using infrared spectroscopy. Note that the difference between the results of the two preceding studies is discussed in detail in \citet{koljonen2017}. 

Figure \ref{fig:mass_constraint} shows planes where the horizontal and vertical axes represent the masses of the WR star and the compact object, respectively. By substituting the orbital velocity amplitude of the compact object obtained in this study $K_{\rm C} = 430^{+150}_{-140}~\si{km.s^{-1}}$ into Equation \ref{eq:mass_WR}, the masses of this system are constrained to the regions shown in red on these planes. The panel (a) shows the case with the assumed inclination angle $i = \ang{25}$, while the panel (b) shows the case with $i = \ang{35}$. The region shown in blue is obtained from the relation 
\begin{equation}
  \frac{2\dot{M}}{M_{\rm tot}} = \frac{\dot{P}}{P} \label{eq:mass_loss_orbital_period_derivative}
\end{equation}
where $\dot{M}$ is the mass loss rate of the system by the stellar wind of the WR star and $\dot{P}$ is the orbital period derivative. This relation is derived from the slowing down of the binary orbit due to the angular momentum loss, with the assumptions that the binary is detached, that any effects from tidal interaction can be ignored (e.g., \cite{bagot1996}), and that the stellar wind removes almost all the specific angular momentum with only a small fraction being accreted onto the compact object. The value of $\dot{P}/P$ has been reported as $\SI{1.02e-6}{yr^{-1}}$ \citep{antokhin2019}. The mass loss rate is estimated in \citet{koljonen2017} to be $\dot{M} = (0.6-2.0)\times \num{e-5}\,M_\odot~\si{yr^{-1}}$ by WR atmospheric modeling of infrared spectra, assuming a wind clumping factor of $D=3.3\text{--}14.3$ \citep{szostek2008} and a distance of $d=\SI{10.2}{kpc}$. An additional constraint on the masses can be obtained from the empirical relation between the mass loss rates and the masses of Galactic WR stars of the WN type, which is given by \citet{zdziarski2013} as 
\begin{gather}
  M_{\rm WR} = \ev{M}\qty(\frac{\dot{M}}{\dot{M}_0})^{1/n} \label{eq:empirical_relation} \\
  \dot{M}_0 = (1.9\pm 0.2)\times\num{e-5}\,M_\odot~\si{yr^{-1}},\quad n=2.93\pm 0.38 \nonumber
\end{gather}
where $\ev{M}=14.7\,M_\odot$ is the geometric average of the masses of WR star samples and $\dot{M}_0$ is the mass loss rate at $M_{\rm WR}=\ev{M}$. This relation is valid only for $M_{\rm WR}<22\,M_\odot$. By eliminating $\dot{M}$ from Equation \ref{eq:mass_loss_orbital_period_derivative} and \ref{eq:empirical_relation}, we obtain
\begin{equation}
  M_{\rm C} = \frac{2\dot{M}_0P}{\ev{M}^n\dot{P}}M_{\rm WR}^n - M_{\rm WR} \label{eq:mass_loss_orbital_period_empirical_WR}
\end{equation}
which is shown in green in Figure \ref{fig:mass_constraint}. The masses of the compact object and the WR star are constrained to be $M_{\rm C} = (1.3\text{--}5.1)\,M_\odot$ and $M_{\rm WR}=(9.3\text{--}12)\,M_\odot$ for $i=\ang{25}$, and $M_{\rm C} = (1.3\text{--}24)\,M_\odot$ and $M_{\rm WR}=(9.3\text{--}16)\,M_\odot$ for $i=\ang{35}$, respectively. These WR star masses are consistent with that estimated in \citet{koljonen2017} from the theoretical mass-luminosity relation of WR stars \citep{grafener2011}. The estimated masses of the compact object here do not allow us to determine whether it is a neutron star or a black hole.

\section{Conclusions}\label{sec:conclusion}
We have presented a high-resolution spectral analysis of the Fe Ly$\alpha$ emission and absorption lines of Cyg X-3 obtained by XRISM/Resolve when the system is in its hypersoft state. The spectra extracted from seven different orbital phase bins are well reproduced by a Doppler shift model of emission lines due to the circular motion of the compact object. Gaussian functions are used to represent the emission and absorption lines with a flexible MCMC method, enabling successful parameter estimation even in spectra where the emission and absorption lines overlap. The compact object is found to have a velocity amplitude of $K_{\rm C} = 430^{+150}_{-140}~\si{km.s^{-1}}$. This value is significantly larger than that obtained by \citet{collaboration2024c} which is primarily determined by the Fe He$\alpha$ lines rather than the Fe Ly$\alpha$ lines, suggesting that the emitting regions of the two lines are spatially and kinematically separated. The absorption lines are naturally interpreted as reflecting the velocity structure of the stellar wind from the WR star. A systematic investigation of various other absorption lines is an important direction for future work, as it may shed light on the detailed velocity structure of the stellar wind.

Based on the result of the velocity amplitude of the compact object, we have discussed the mass of the compact object and the WR star. The mass function obtained by substituting the velocity amplitude derived in this study, combined with the relation between the mass loss rate and the orbital period derivative and the empirical relation of the mass loss rate and the mass of the Galactic WN type WR stars, allows us to constrain the mass of the compact object to be $M_{\rm C} = (1.3\text{--}5.1)\,M_\odot$ and $M_{\rm WR}=(9.3\text{--}12)\,M_\odot$ for the assumed orbital inclination angle of $i=\ang{25}$, and $M_{\rm C} = (1.3\text{--}24)\,M_\odot$ and $M_{\rm WR}=(9.3\text{--}16)\,M_\odot$ for $i=\ang{35}$, respectively. This means the identity of the compact object is still uncertain.

\section*{Acknowledgments}
The authors appreciate Hajime Kawahara and Shota Miyazaki at ISAS for their valuable advice and discussions on the data analysis method. We also express our gratitude to Elisa Costantini at SRON for her useful comments on the manuscript. We thank all those who contributed to the XRISM mission.

\appendix

\section{Additional Spectral Fits and Parameters}\label{sec:appendix_additional}
\begin{figure*}
  \begin{tabular}{cccc}
    \begin{minipage}{0.25\linewidth}
      \centering
      \includegraphics[width=\linewidth]{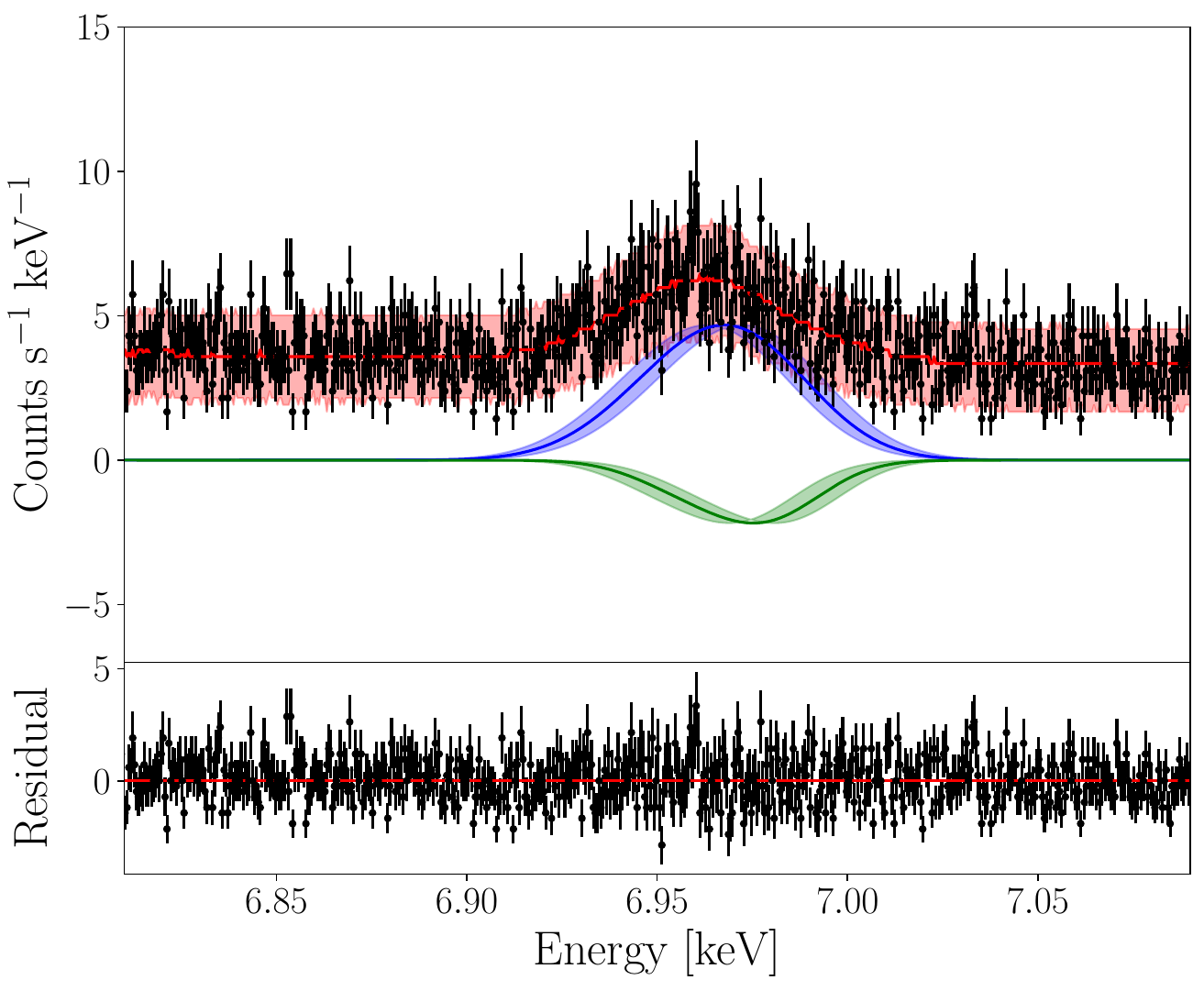}
      \subcaption{$\varphi=0.000\text{--}0.125$}
    \end{minipage}
    \begin{minipage}{0.25\linewidth}
      \centering
      \includegraphics[width=\linewidth]{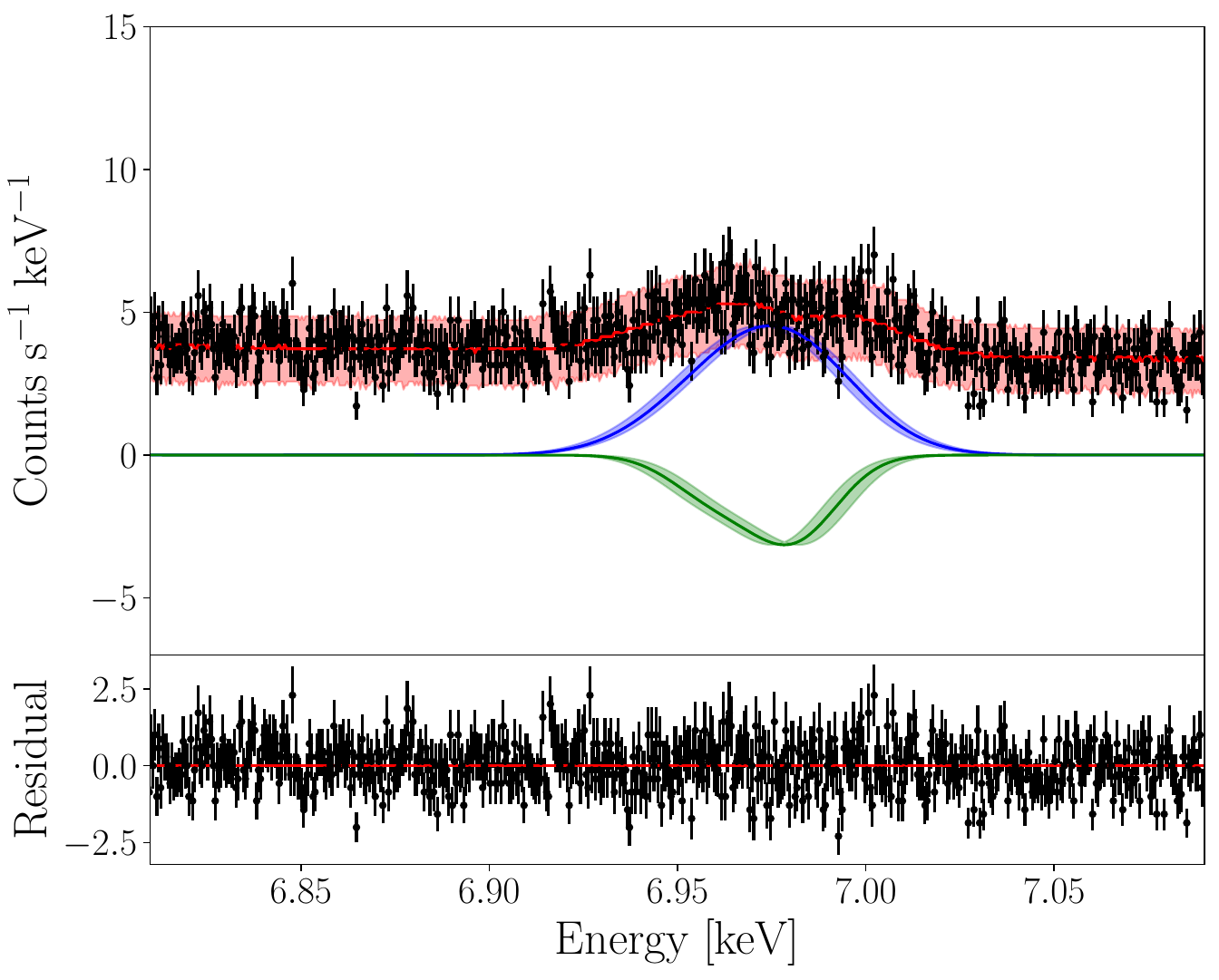}
      \subcaption{$\varphi=0.125\text{--}0.250$}
    \end{minipage}
    \begin{minipage}{0.25\linewidth}
      \centering
      \includegraphics[width=\linewidth]{spec_result_bin3.pdf}
      \subcaption{$\varphi=0.375\text{--}0.500$}
    \end{minipage}
    \begin{minipage}{0.25\linewidth}
      \centering
      \includegraphics[width=\linewidth]{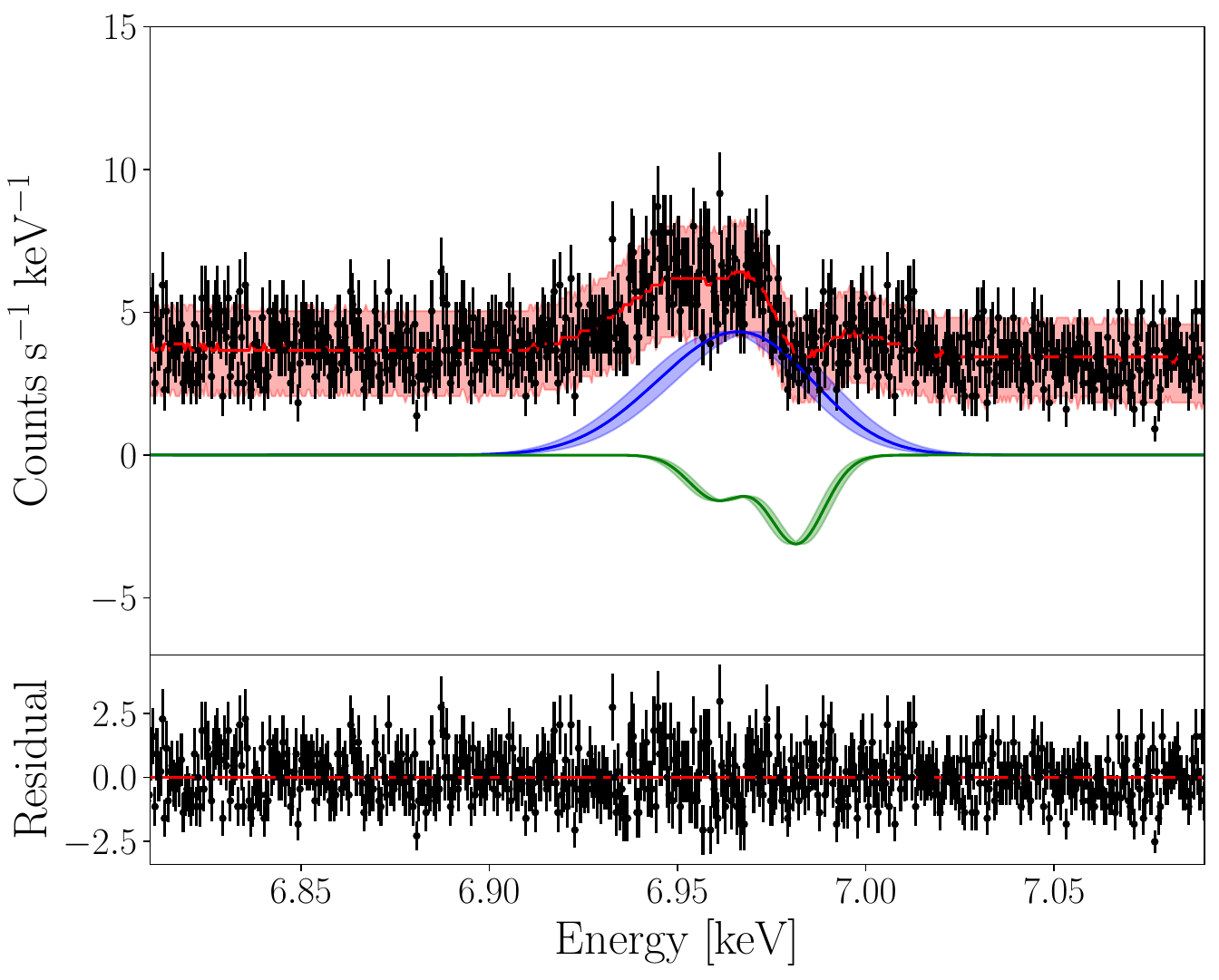}
      \subcaption{$\varphi=0.500\text{--}0.625$}
    \end{minipage}\\
    \begin{minipage}{0.25\linewidth}
      \centering
      \includegraphics[width=\linewidth]{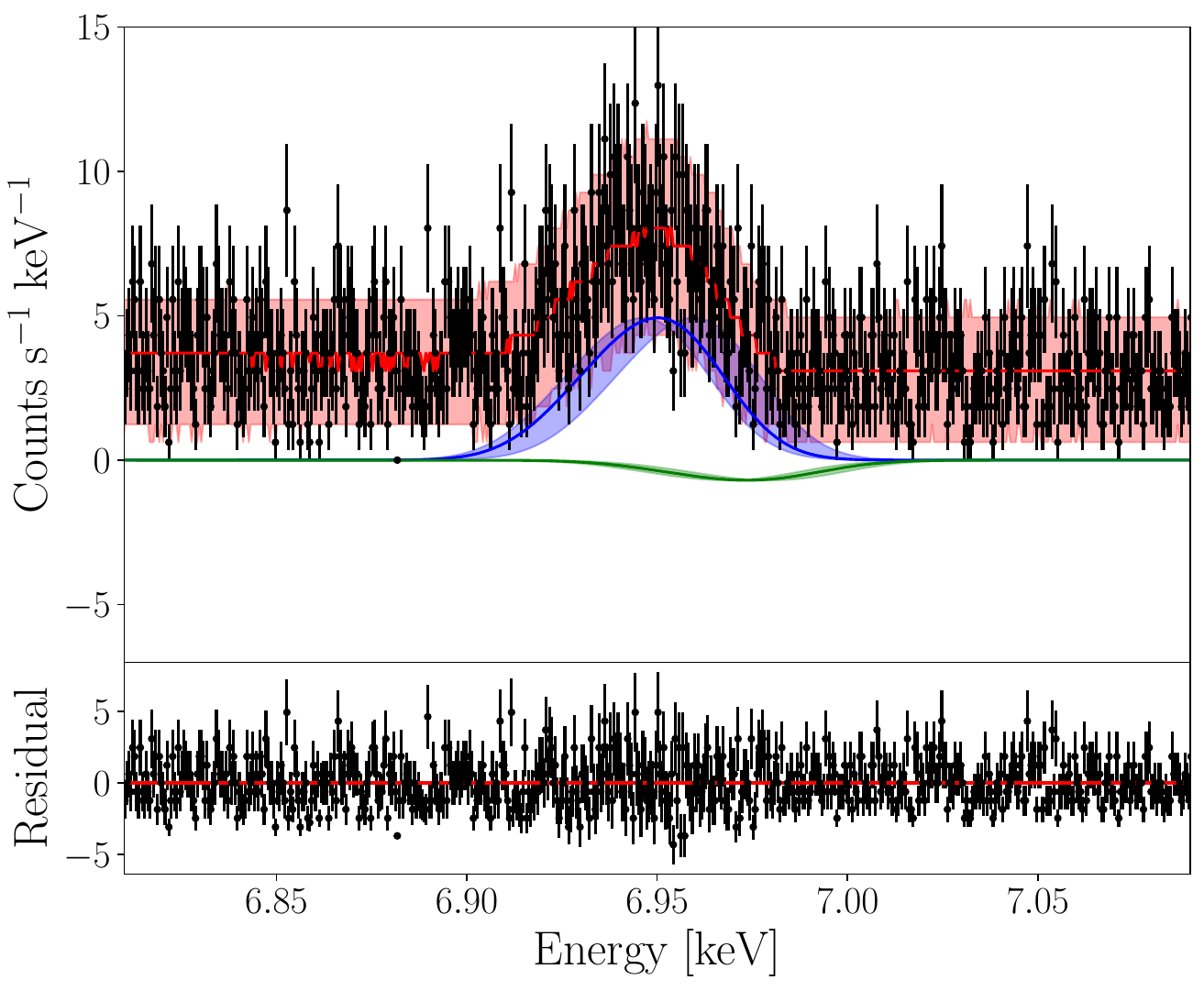}
      \subcaption{$\varphi=0.625\text{--}0.750$}
    \end{minipage}
    \begin{minipage}{0.25\linewidth}
      \centering
      \includegraphics[width=\linewidth]{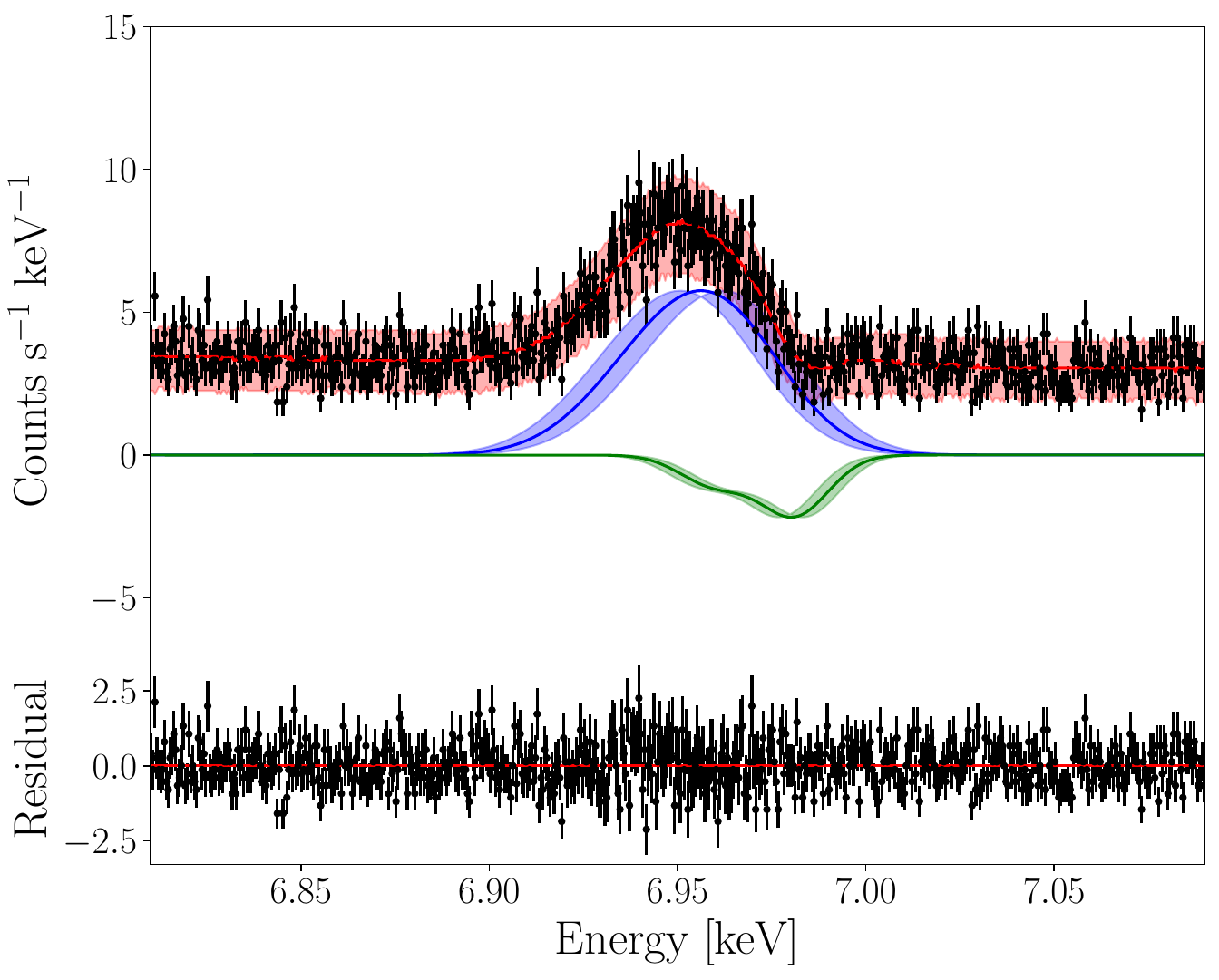}
      \subcaption{$\varphi=0.750\text{--}0.875$}
    \end{minipage}
    \begin{minipage}{0.25\linewidth}
      \centering
      \includegraphics[width=\linewidth]{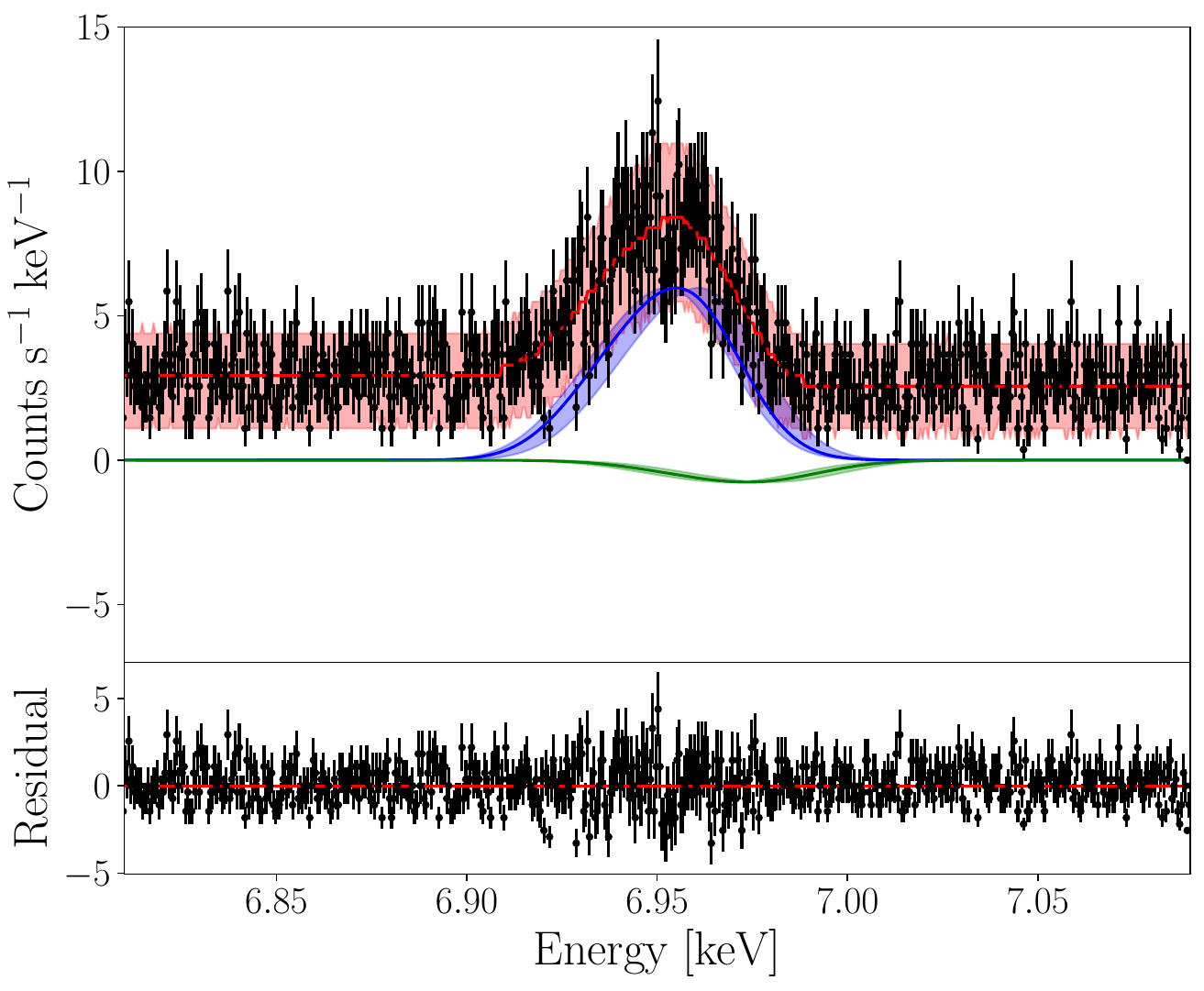}
      \subcaption{$\varphi=0.875\text{--}1.000$}
    \end{minipage}
  \end{tabular}
  \caption{
    The predictions (median) with the $90\%$ HPDIs from the fitting for seven different phase bins overlaid on the observed spectra for the analysis of each phase independently(subsection \ref{sec:analysis_ind}). For each phase bin, the top panel shows the predicted median spectrum (red solid line) and the $90\%$ HPDI (red-shaded region). The blue and green solid lines show the emission and absorption lines with the medians of posterior samples of the parameters described in Table \ref{tab:prior}. The blue- and green-shaded regions correspond to the $90\%$ HPDIs for redshifts ($z_{\rm e}$ and $z_{\rm a}$). The bottom panel shows the residual from the median.
  }
  \label{fig:spec_ind}
\end{figure*}

Figure \ref{fig:spec_ind} shows the predictions (median) with the $90\%$ HPDIs from the fitting for all seven phases overlaid on the observed spectra for the analysis of each phase independently. Table \ref{tab:median_hpdi_ind} provides the medians of the marginal posteriors and their $90\%$ HPDIs with the $C$-statistic/dofs. 

\begin{figure*}
  \begin{tabular}{cccc}
    \begin{minipage}{0.25\textwidth}
      \centering
      \includegraphics[width=\linewidth]{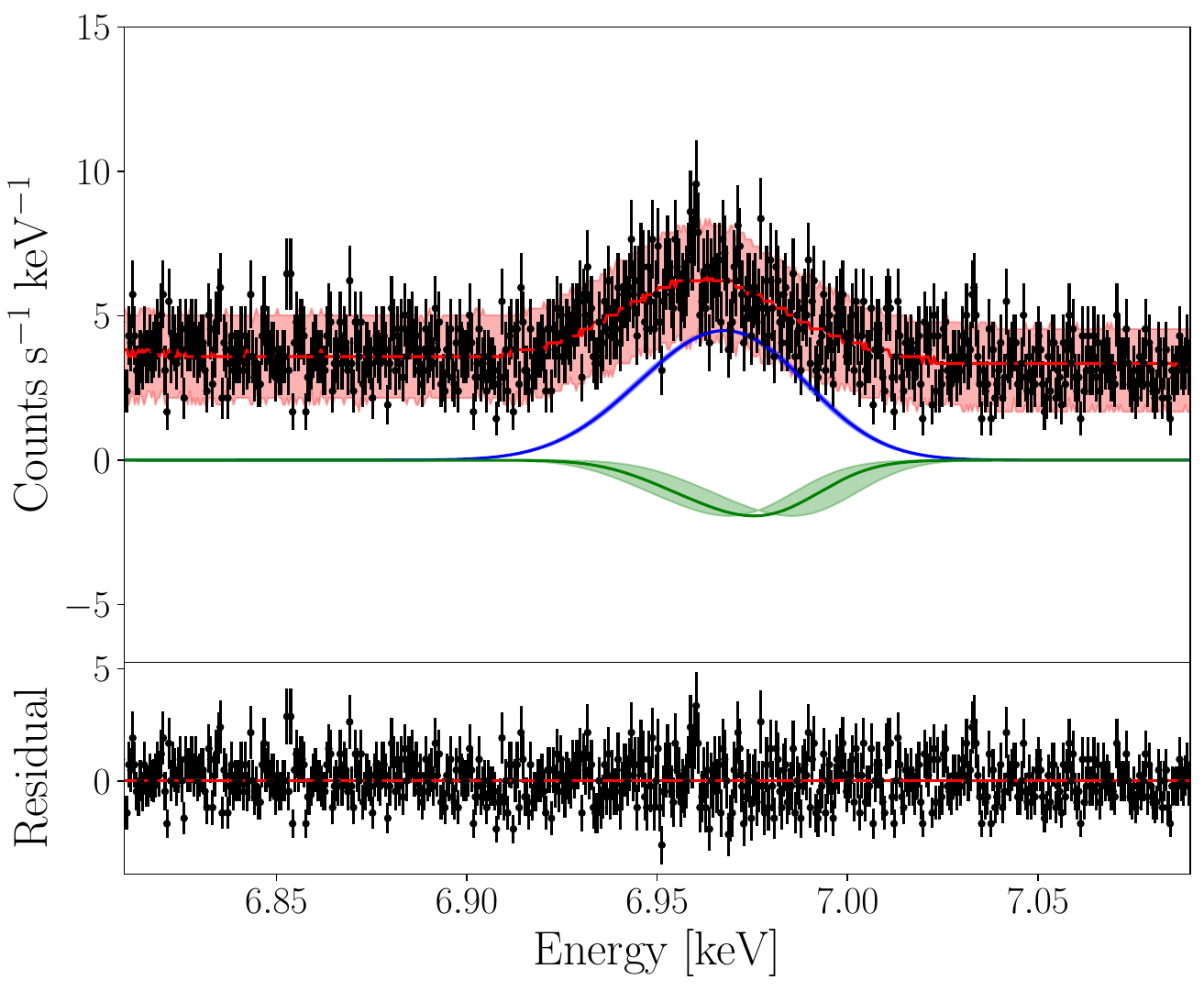}
      \subcaption{$\varphi=0.000\text{--}0.125$}
    \end{minipage}
    \begin{minipage}{0.25\textwidth}
      \centering
      \includegraphics[width=\linewidth]{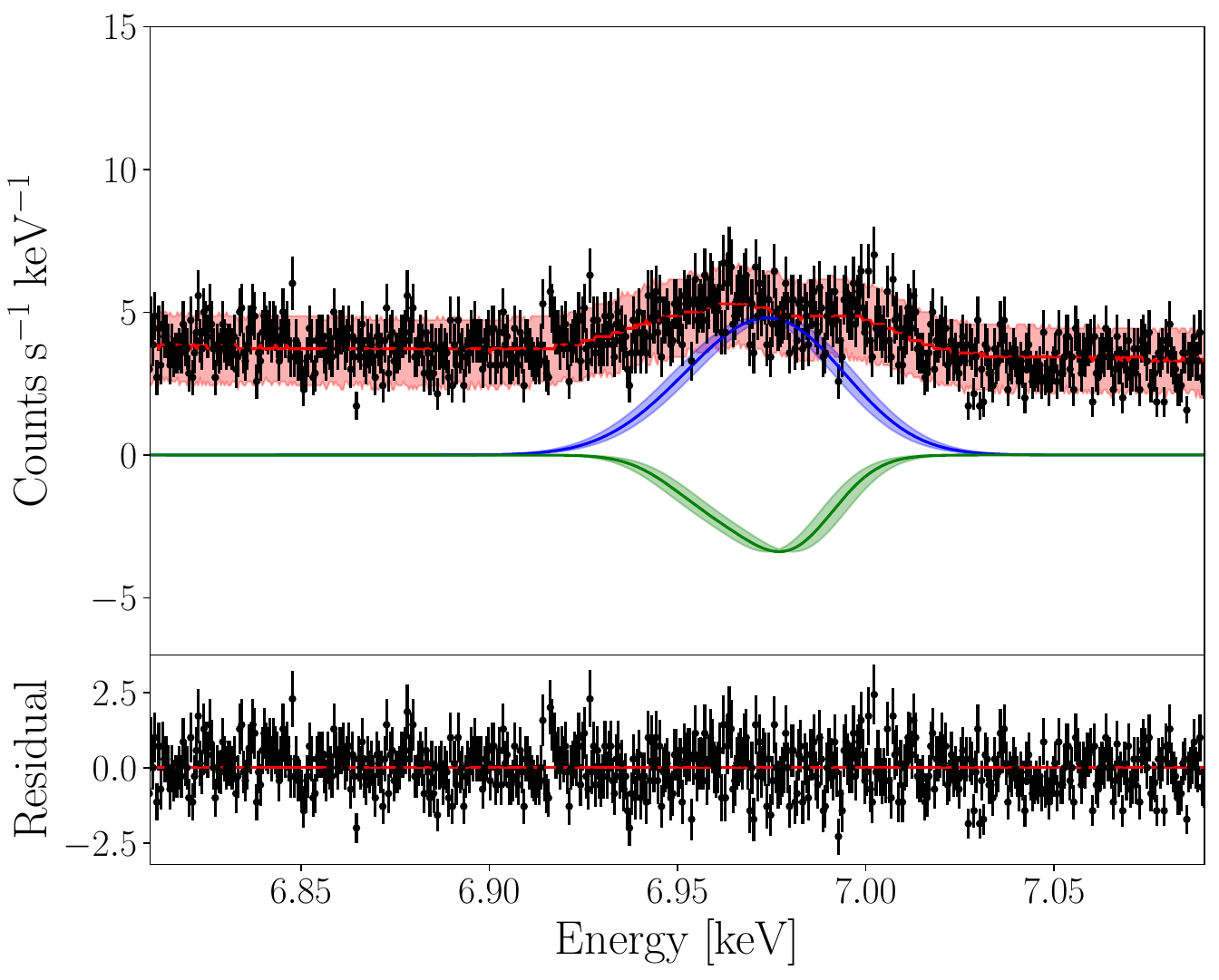}
      \subcaption{$\varphi=0.125\text{--}0.250$}
    \end{minipage}
    \begin{minipage}{0.25\textwidth}
      \centering
      \includegraphics[width=\linewidth]{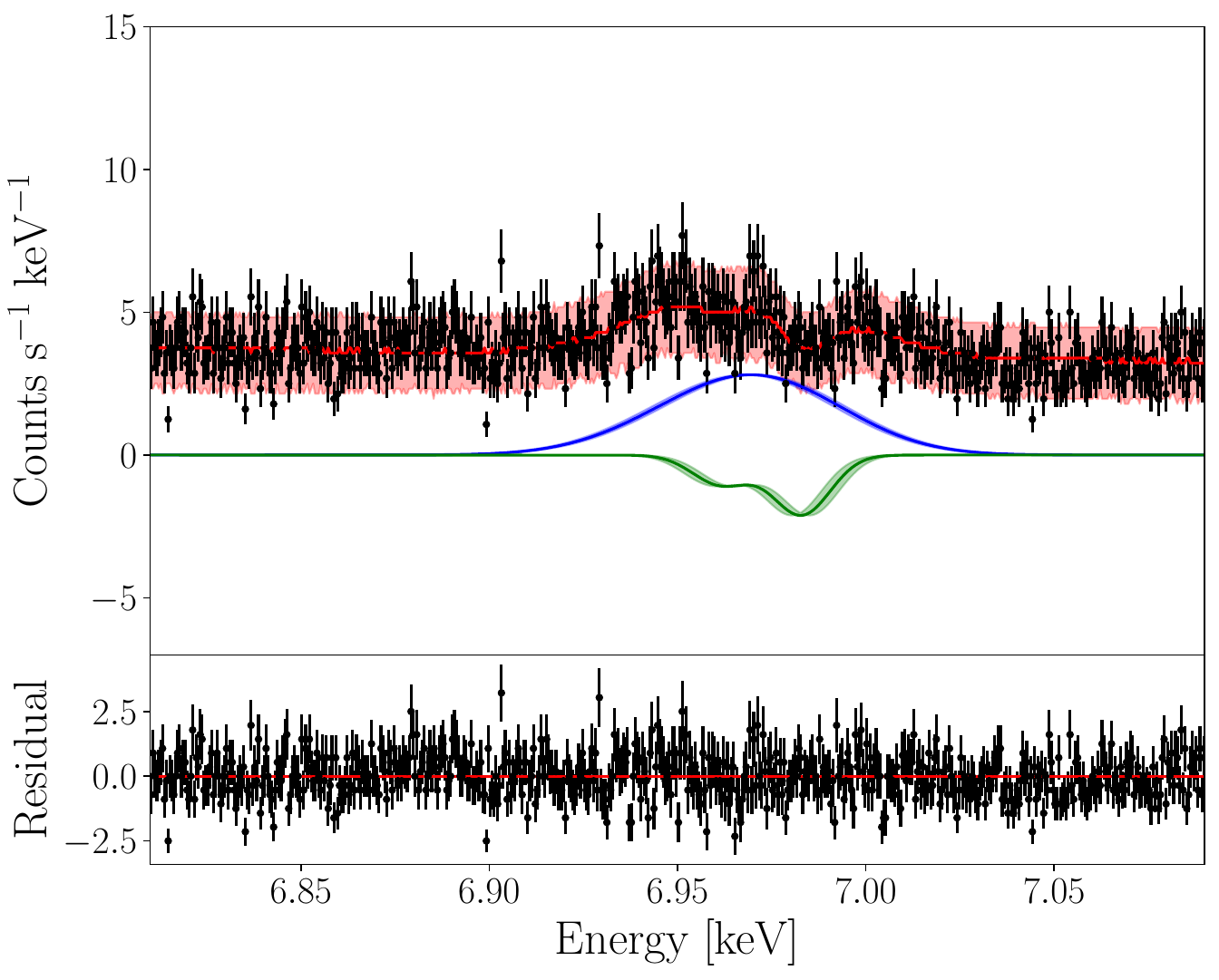}
      \subcaption{$\varphi=0.375\text{--}0.500$}
    \end{minipage}
    \begin{minipage}{0.25\textwidth}
      \centering
      \includegraphics[width=\linewidth]{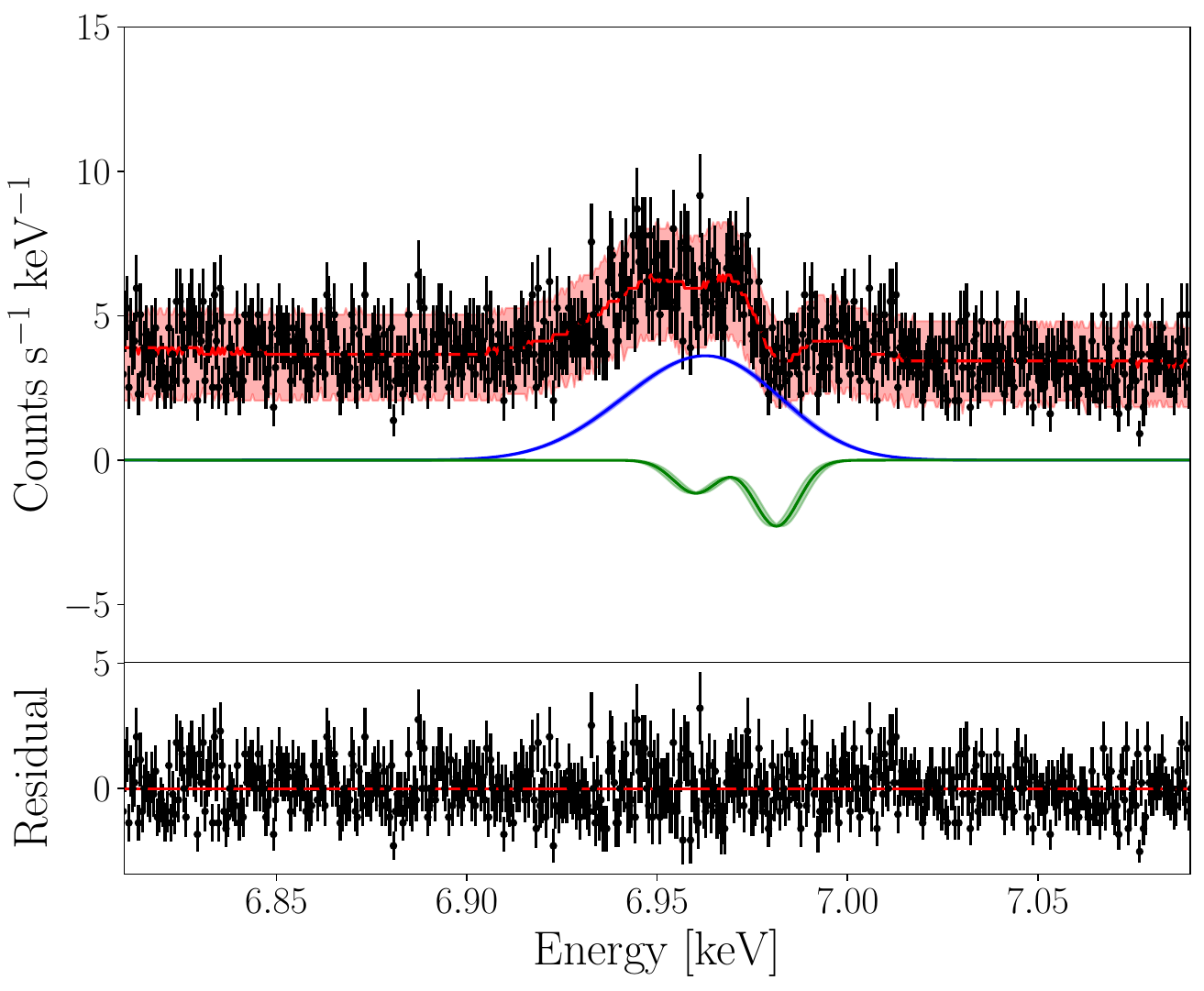}
      \subcaption{$\varphi=0.500\text{--}0.625$}
    \end{minipage}\\
    \begin{minipage}{0.25\linewidth}
      \centering
      \includegraphics[width=\linewidth]{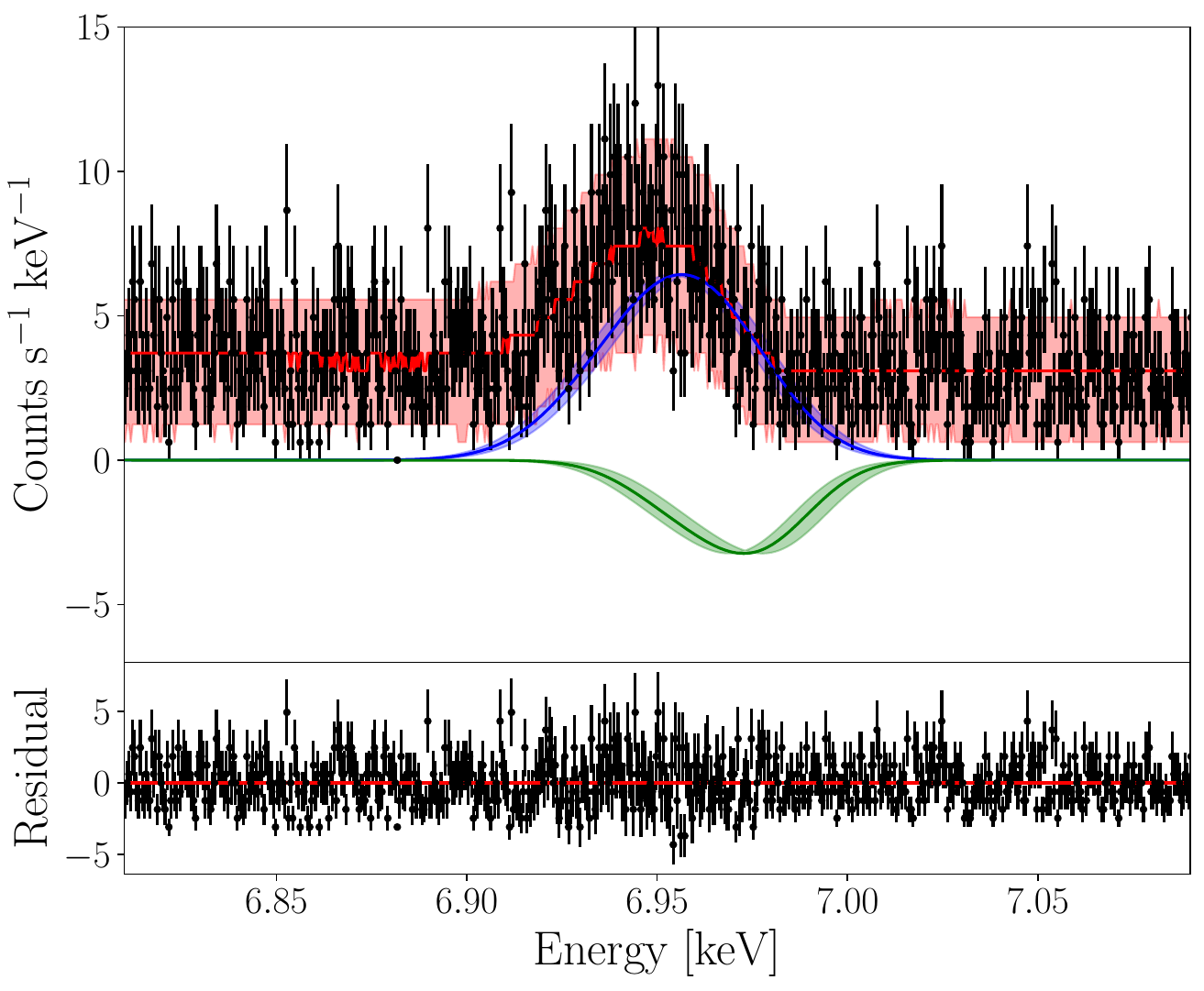}
      \subcaption{$\varphi=0.625\text{--}0.750$}
    \end{minipage}
    \begin{minipage}{0.25\linewidth}
      \centering
      \includegraphics[width=\linewidth]{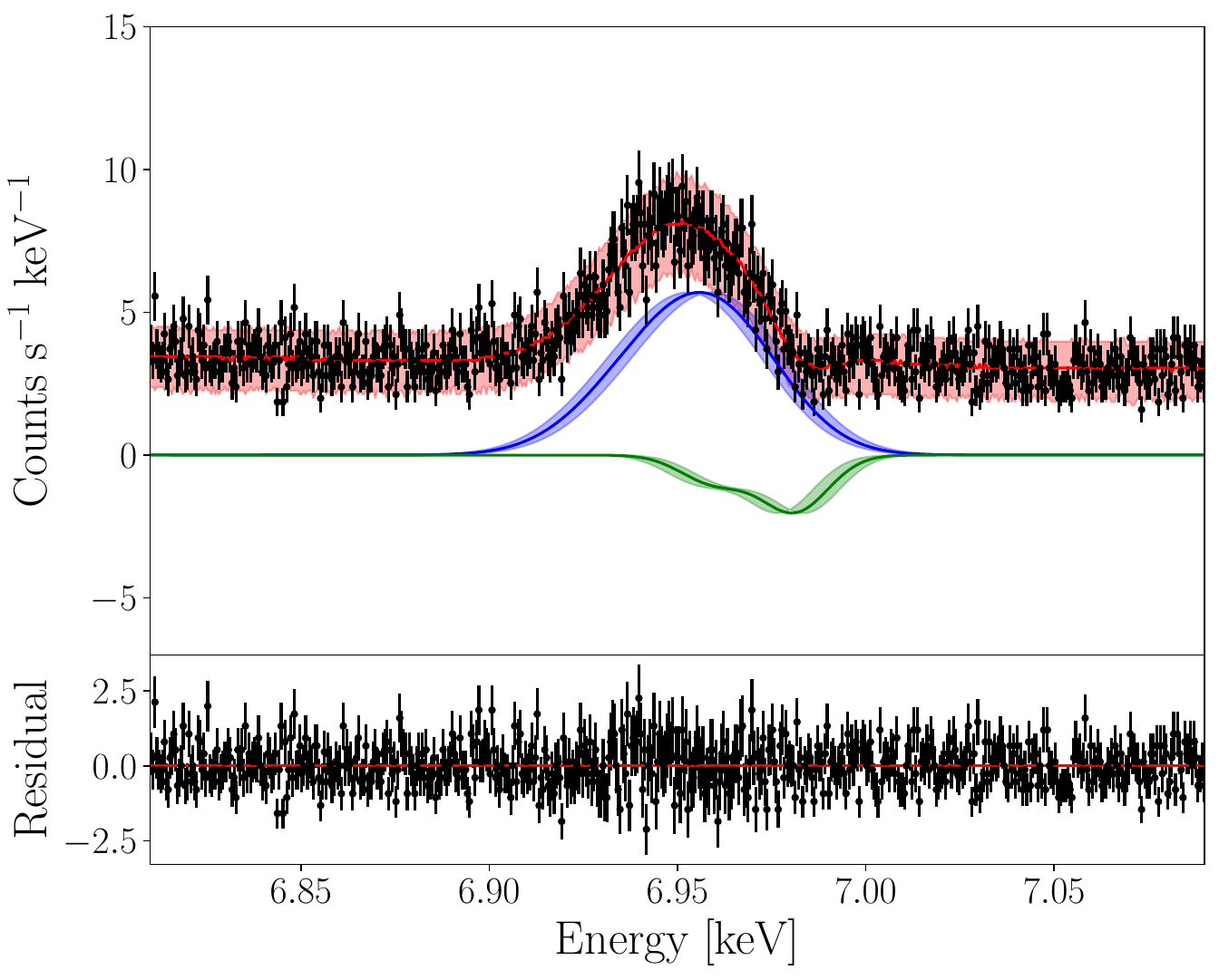}
      \subcaption{$\varphi=0.750\text{--}0.875$}
    \end{minipage}
    \begin{minipage}{0.25\linewidth}
      \centering
      \includegraphics[width=\linewidth]{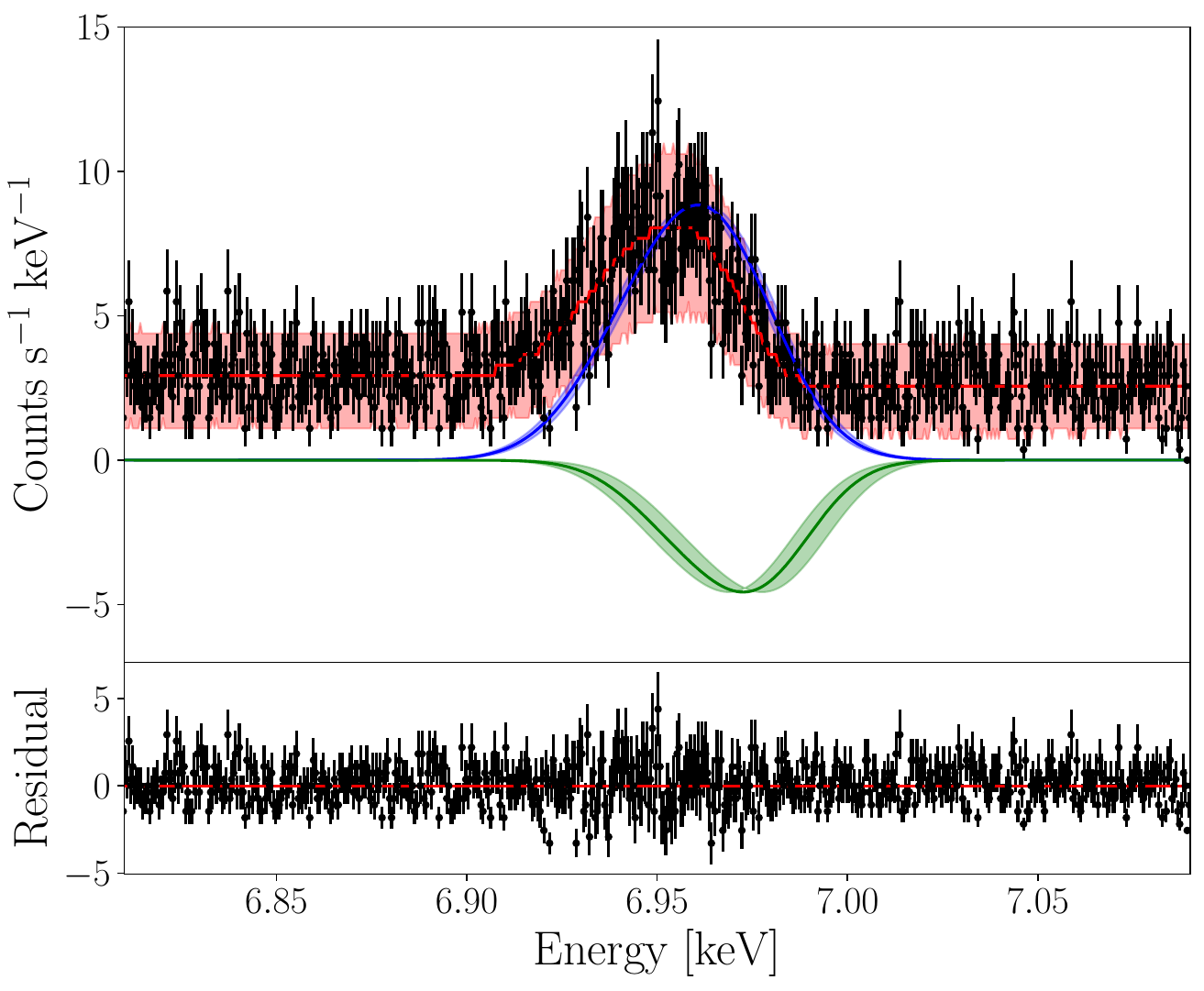}
      \subcaption{$\varphi=0.875\text{--}1.000$}
    \end{minipage}
  \end{tabular}
  \caption{
    The same figure as Figure \ref{fig:spec_ind} but obtained by the analysis assuming a circular orbit(subsection \ref{sec:analysis_sine}).
  }
  \label{fig:spec_sine}
\end{figure*}

Figure \ref{fig:spec_sine} shows the predictions (median) with the $90\%$ HPDIs from the fitting for all seven phases overlaid on the observed spectra for the analysis assuming a circular orbit. Table \ref{tab:median_hpdi_sine} provides the medians of the marginal posteriors and their $90\%$ HPDIs with the $C$-statistic/dof.

\begin{table*}[h]
  \caption{Median and 90\% interval of inferred parameters when analyzing each phase independently(subsection \ref{sec:analysis_ind}).}
  \label{tab:median_hpdi_ind}
  \centering
  \begin{threeparttable}
    \begin{tabularx}{\linewidth}{lccccccc}\hline\hline\noalign{\vskip3pt}
      Parameter~\textbackslash~Phase\tnote{*} & $\varphi = 0.0625$ & $\varphi=0.1875$ & $\varphi=0.4375$ & $\varphi=0.5625$ & $\varphi=0.6875$ & $\varphi=0.8125$ & $\varphi=0.9375$  \\ \hline\noalign{\vskip3pt}
      \multicolumn{8}{c}{Continuum} \\ \hline\noalign{\vskip3pt}
      Flux ($\SI{e-9}{erg.cm^{-2}.s^{-1}}$)\tnote{**} & $1.17^{+0.03}_{-0.02}$ & $1.19 \pm 0.02$ & $1.17\pm 0.02$ & $1.20^{+0.03}_{-0.02}$ & $1.12 \pm 0.04$ & $1.07 \pm 0.02$ & $0.94\pm 0.03$ \\
      Photon index ($\alpha$)& $2.73\pm 0.08$ & $2.73^{+0.07}_{-0.09}$ & $2.73 \pm 0.08$ & $2.72^{+0.08}_{-0.09}$ & $2.76^{+0.08}_{-0.09}$ & $2.77^{+0.09}_{-0.08}$ & $2.83^{+0.08}_{-0.09}$ \\ \hline\noalign{\vskip3pt}
      \multicolumn{8}{c}{Emission lines} \\ \hline\noalign{\vskip3pt}
      Norm ($N_{\rm e}$)~$\times \num{e-3}$ & $1.2^{+1.0}_{-0.5}$ & $1.2^{+0.8}_{-0.6}$ & $0.8\pm 0.2$ & $1.1^{+0.5}_{-0.4}$ & $1.1 \pm 0.2$ & $1.4^{+0.6}_{-0.4}$ & $1.3^{+0.7}_{-0.2}$  \\
      Width ($\sigma_{\rm e}$) [$\si{keV}$] ~$\times \num{e-2}$& $1.9 \pm 0.3$ & $1.8^{+0.5}_{-0.3}$ & $2.2^{+0.4}_{-0.3}$ & $1.7^{+0.2}_{-0.3}$ & $1.5^{+0.5}_{-0.3}$ & $1.7^{+0.2}_{-0.3}$ & $1.4^{+0.3}_{-0.2}$ \\
      Redshift ($z_{\rm e}$)~$\times \num{e-3}$ & $-0.0\pm 0.7$ & $-1.0^{+0.4}_{-0.3}$ & $-0.0^{+0.6}_{-0.5}$ & $0.0^{+0.6}_{-0.7}$ & $2.6^{+0.6}_{-0.1}$ & $1.6\pm{0.8}$ & $2.0^{+0.4}_{-0.9}$ \\ \hline\noalign{\vskip3pt}
      \multicolumn{8}{c}{Absorption lines} \\ \hline\noalign{\vskip3pt}
      Norm ($N_{\rm a}$)~$\times \num{e-3}$& $-0.5^{+0.5}_{-1.0}$ & $-0.6^{+0.6}_{-0.8}$ & $-0.2\pm 0.2$ & $-0.4^{+0.3}_{-0.5}$ & $-0.2^{+0.2}_{-0.9}$ & $-0.4^{+0.3}_{-0.6}$ & $-0.2^{+0.2}_{-0.6}$ \\
      Width ($\sigma_{\rm a}$) [$\si{keV}$] ~$\times \num{e-2}$& $1.4^{+0.9}_{-0.6}$ & $1.1^{+0.3}_{-0.4}$ & $0.6^{+0.4}_{-0.2}$ & $0.7 \pm 0.3$ & $1.6^{+1.4}_{-0.7}$ & $0.9^{+0.5}_{-0.4}$ & $1.6^{+1.1}_{-0.8}$ \\
      Redshift ($z_{\rm a}$)~$\times \num{e-3}$ & $-0.9^{+0.9}_{-0.8}$ & $-1.1^{+0.5}_{-0.6}$ & $-1.4\pm 0.3$ & $-1.2\pm 0.3$ & $-0.8^{+0.8}_{-0.9}$ & $-1.1^{+0.5}_{-0.4}$ & $-0.8^{+0.8}_{-0.9}$ \\ \hline\noalign{\vskip3pt}
      $C$-statistic / dof\tnote{\textdagger} & 620 / 592 & 622 / 592 & 641 / 592 & 598 / 592 & 635 / 592 & 586 / 592 & 628 / 592 \\ \hline\noalign{\vskip3pt} 
      \end{tabularx}
      \begin{tablenotes}
        \item[*] Each orbital phase $\varphi$ is shown as the median of the phase range from which each spectrum is extracted.
        \item[**] The flux in $6.0-\SI{8.0}{keV}$.
        \item[\textdagger] These values were calculated from the medians of the model predictions.
      \end{tablenotes}
  \end{threeparttable}
\end{table*}

\begin{table*}
  \caption{Median and 90\% interval of inferred parameters obtained by the analysis assuming a circular orbit(subsection \ref{sec:analysis_sine}).}
  \label{tab:median_hpdi_sine}
  \centering
  \begin{threeparttable}
    \begin{tabularx}{\linewidth}{lccccccc}\hline\hline\noalign{\vskip3pt}
      Parameter~\textbackslash~Phase\tnote{*} & $\varphi = 0.0625$ & $\varphi=0.1875$ & $\varphi=0.4375$ & $\varphi=0.5625$ & $\varphi=0.6875$ & $\varphi=0.8125$ & $\varphi=0.9375$  \\ \hline\noalign{\vskip3pt}
      \multicolumn{8}{c}{Continuum} \\ \hline\noalign{\vskip3pt}
      Flux ($\SI{e-9}{erg.cm^{-2}.s^{-1}}$)\tnote{**} & $1.17^{+0.02}_{-0.03}$ & $1.19 \pm 0.02$ & $1.17 \pm 0.02$ & $1.20^{+0.03}_{-0.02}$ & $1.12\pm 0.04$ & $1.07 \pm 0.02$ & $0.94\pm 0.03$ \\
      Photon index ($\alpha$)& $2.73^{+0.08}_{-0.09}$ & $2.73\pm 0.08$ & $2.73 \pm 0.08$ & $2.72^{+0.08}_{-0.09}$ & $2.75 \pm 0.08$ & $2.77 \pm 0.08$ & $2.83^{+0.08}_{-0.09}$ \\ \hline\noalign{\vskip3pt}
      \multicolumn{8}{c}{Emission lines} \\ \hline\noalign{\vskip3pt}
      Norm ($N_{\rm e}$)~$\times \num{e-3}$ & $1.2^{+1.5}_{-0.5}$ & $1.2^{+1.8}_{-0.7}$ & $0.8^{+0.4}_{-0.2}$ & $0.9 \pm 0.2$ & $1.7^{+0.9}_{-0.6}$ & $1.4^{+0.5}_{-0.4}$ & $2.1^{+1.6}_{-0.9}$ \\
      Width ($\sigma_{\rm e}$) [$\si{keV}$] ~$\times \num{e-2}$ & $1.9 \pm 0.4$ & $1.8^{+0.5}_{-0.4}$ & $2.1^{+0.3}_{-0.4}$ & $1.7 \pm 0.2$ & $1.8 \pm 0.3$ & $1.7 \pm 0.3$ & $1.6^{+0.3}_{-0.2}$ \\
      Velocity amp. ($v$) [$\si{km.s^{-1}}$] & \multicolumn{7}{c}{$430^{+150}_{-140}$} \\ 
      Velocity offset ($v_0$) [$\si{km.s^{-1}}$] & \multicolumn{7}{c}{$100^{+93}_{-110}$} \\
      Phase offset ($\varphi_0$) & \multicolumn{7}{c}{$-0.02^{+0.05}_{-0.04}$} \\ \hline\noalign{\vskip3pt}
      \multicolumn{8}{c}{Absorption lines} \\ \hline\noalign{\vskip3pt}
      Norm ($N_{\rm a}$)~$\times \num{e-3}$& $-0.4^{+0.4}_{-0.2}$ & $-0.7^{+0.7}_{-1.8}$ & $-0.3^{+0.2}_{-0.5}$ & $-0.3^{+0.1}_{-0.2}$ & $-0.7^{+0.5}_{-0.8}$ & $-0.3^{+0.3}_{-0.5}$ & $-1.0^{+0.8}_{-1.6}$ \\
      Width ($\sigma_{\rm a}$) [$\si{keV}$] ~$\times \num{e-2}$& $1.4^{+1.0}_{-0.7}$ & $1.2 \pm 0.4$ & $0.7^{+0.5}_{-0.3}$ & $0.6\pm 0.2$ & $1.4\pm 0.4$ & $0.9^{+0.4}_{-0.5}$ & $1.4\pm 0.3$ \\
      Redshift ($z_{\rm a}$)~$\times \num{e-3}$ & $-1.0^{+1.0}_{-1.4}$ & $-1.0\pm 0.5$ & $-1.4 \pm 0.3$ & $-1.2\pm 0.2$ & $-0.6^{+0.6}_{-0.7}$ & $-1.1^{+0.6}_{-0.5}$ & $-0.6^{+0.6}_{-0.7}$ \\ \hline\noalign{\vskip3pt}
      $C$-statistic / dof\tnote{\textdagger} & \multicolumn{7}{c}{4318 / 4148} \\ \hline\noalign{\vskip3pt}
      \end{tabularx}
      \begin{tablenotes}
        \item[*] Each orbital phase $\varphi$ is shown as the median of the phase range from which each spectrum is extracted.
        \item[**] The flux in $6.0-\SI{8.0}{keV}$.
        \item[\textdagger] This value was calculated from the median of the model predictions.
      \end{tablenotes}
  \end{threeparttable}
\end{table*}

\bibliography{cyg_x3_pasj,collaboration}
\bibliographystyle{aasjournal}

\end{document}